\documentclass[structabstract]{aa}  
\usepackage{amsmath}
\usepackage{graphicx}
\usepackage{txfonts}
\usepackage{amssymb}
\usepackage[]{natbib}
\usepackage{mathrsfs}
\usepackage[colorlinks=true, linkcolor=blue, citecolor=blue, urlcolor=blue]{hyperref}
\usepackage{caption}

\newcommand{\qt}[0]{\mbox{Qatar-1}}
\newcommand{\qta}[0]{\mbox{Qatar-1A}}
\newcommand{\qtb}[0]{\mbox{Qatar-1b}}

\newcommand{\bjdtdb}[0]{\mbox{$\mathrm{BJD}_{\mathrm{TDB}}$}}

\newcommand{\hjdutc}[0]{\mbox{$\mathrm{HJD}_{\mathrm{UTC}}$}}
\begin{document}

   \title{Qatar-1: indications for possible transit timing variations
     \thanks{Full Tables of all the primary transits are only available in electronic form at the CDS via http://cdsweb.u-strasbg.fr/cgi-bin/qcat?J/A+A/.}}
   \subtitle{}

   \author{C. von Essen$^{1}$, S. Schr\"oter$^1$, E. Agol$^{2}$ \and J.H.M.M. Schmitt$^{1}$
          }
   \authorrunning{C. von Essen et al.}
   \titlerunning{Transit timing variations on Qatar-1}
   \offprints{cessen@hs.uni-hamburg.de}

   \institute{$^1$Hamburger Sternwarte, University of Hamburg,
              Gojenbergsweg 112, 21029 Hamburg, Germany\\
	      $^2$Dept. of Astronomy, Box 351580, University of Washington, 
              Seattle, WA 98195\\
              \email{cessen@hs.uni-hamburg.de}
               }

   \date{Received XXXX; accepted XXXX}

\abstract
 {}
  {Variations in the timing of transiting exoplanets provide a
    powerful tool detecting additional planets in the system. Thus,
    the aim of this paper is to discuss the plausibility of transit
    timing variations on the \qt\ system by means of primary transit
    light curves analysis. Furthermore, we provide an interpretation
    of the timing variation.}
  {We observed \qt\ between March 2011 and October 2012 using the
    1.2~m OLT telescope in Germany and the 0.6~m PTST telescope in
    Spain. We present 26 primary transits of the hot Jupiter Qatar-1b.
    In total, our light curves cover a baseline of 18 months.}
  {We report on indications for possible long-term transit timing
    variations (TTVs). Assuming that these TTVs are true, we present
    two different scenarios that could explain them. Our reported
    $\sim$ 190 days TTV signal can be reproduced by either a weak
    perturber in resonance with \qtb, or by a massive body in the
    brown dwarf regime. More observations and radial velocity
    monitoring are required to better constrain the perturber's
    characteristics. We also refine the ephemeris of Qatar-1b, which
    we find to be \mbox{$T_0 = 2456157.42204 \pm 0.0001$~\bjdtdb} and
    \mbox{$P = 1.4200246 \pm 0.0000007$~days}, and improve the system
    orbital parameters.}
  {}


   \maketitle
%

\section{Introduction}

The discovery of Neptune in 1846 was a milestone in astronomy. The
position at which the planet was detected by J. G. Galle had been
predicted -- independently -- by U.J. Le Verrier and J. C. Adams, who
attributed the observed irregularities of Uranus' orbit to the
gravitational attraction of another perturbing body outside Uranus'
radius. Thus, Neptune became the first planet to be predicted by
celestial mechanics before it was directly observed.

Similar to the case of Neptune, in the realm of exoplanets an
additional planet or exomoon can reveal itself by its gravitational
influence on the orbital elements of the observed planet. Most
notably, for transiting planets a perturber is expected to induce
short-term transit timing variations (TTVs)
(\cite{Holman2005,Agol2005}); because timing measurements can be
carried out quite accurately, the TTV searches are quite sensitive,
and it is possible to detect additional objects in the stellar system
and to derive their properties with TTVs.

The detection of TTVs in ground-based measurements requires both a
sufficiently long baseline and good phase coverage. The search for
planets by TTVs has been a major activity in ground-based exoplanet
research. So far, TTVs have been claimed in WASP-3b
\citep{Maciejewski2010}, WASP-10b \citep{Maciejewski2011}, WASP-5b
\citep{Fukui2011}, HAT-P-13 (\cite{Pal2011}, \cite{Nascimbeni2011},
but see \cite{Fulton2011}), and OGLE-111b (\cite{Diaz2008}, but see
\cite{Adams2010}).

In recent years, the search for TTVs in extrasolar planetary systems
has entered a new era, marked by the advent of the space-based
observatories CoRoT and {\it Kepler}, which provide photometry of
unprecedented accuracy. On one hand, the {\it Kepler} team has already
presented more than 40 TTVs in exoplanetary systems, such as Kepler-9
(\cite{Holman2010}), Kepler-11 \citep{Lissauer2011}, Kepler-19
\citet{2011ApJ...743..200B}, Kepler-18 \citep{Cochran2011}, and Kepler
29, 30, 31 and 32 \citep{Fabrycky2012}.  On the other hand,
\cite{Steffen2012} searched for planetary companions orbiting
hot-Jupiter planet candidates and found that most of the systems show
no significant TTV signal.  Yet, the {\it Kepler} satellite observes
only a small area of the sky and has a limited lifetime, thus TTV will
continue to play a major role also in ground-based studies.

The transiting Jovian planet \qtb \ was discovered by
\cite{Alsubai2011} as the first planet found within The Qatar
Exoplanet Survey. \qta \ is an old (\mbox{$> 4$~Gy}) K3V~star with
\mbox{0.85~M$_\odot$} and \mbox{0.82~R$_\odot$}, orbited by a close-in
(\mbox{$0.023$~AU}) planet with a mass of \mbox{$1.1$~M$_J$} and a
period of \mbox{$1.42$~d}. \qtb \ has a radius of \mbox{$1.16$~R$_J$}
and an inclination angle of $83.47^\circ$, implying a nearly grazing
transit. Therefore \qt \ offers an outstanding opportunity to search
for TTVs for several reasons: Its large inclination and close-in orbit
yield a high-impact parameter, making the shape and timing of the
transit sensitive to variations of the orbital parameters. Its large
radius and short orbital period allow for the study of a large number
of deep transits over hundreds of epochs.

We have therefore observed transits in \qtb \ over the past few years,
which we describe here. We specifically describe the observational
setup in section~\ref{sec:ObsRed} as well as the data reduction
process. We subsequently present the details of our light curve
analysis and transit modeling in Sect.~\ref{sec:Analysis} and describe
our TTV analysis in section~\ref{sec:TTV}.  We interpret our results
in Section~\ref{sec:dyn}, where we present different dynamical
scenarios that were tested to suggest the existence of an additional
body in the system. In Section \ref{sec:concl} we conclude.

\section{Observations and data reduction}
\label{sec:ObsRed}

Our observations comprise 18~transits of \qt \ obtained using the
1.2~m Oskar-L\"uhning telescope (OLT) at Hamburg Observatory, Germany,
and 8~transits using the 0.6~m Planet Transit Study Telescope (PTST)
at the Mallorca Observatory in Spain.

The OLT data were taken between March~2011 and October~2012 using an
Apogee Alta~U9000 CCD with a $9' \times 9'$ field of view. Binning was
usually $4\times4$, but the first transit observations were taken in a
$2\times2$ configuration. The binning was increased to reduce
individual exposure times. The data were obtained with typical
exposure times between 45~and 300~seconds depending on the night
quality, the binning configuration, and the star's altitude. All
exposures were obtained using a Johnson-Cousins Schuler
R~filter. Because \qt \ is circumpolar at Hamburg's latitude, typical
airmass values range from 1~up to 1.9, while the average seeing value
is \mbox{2.5~arcsec}.

The PTST data were taken between May and August~2012 using a Santa
Barbara CCD with a $30' \times 30'$ field of view in a $3\times 3$
binning and a Baade~R-band filter setup. The exposure times range from
50~to 90~seconds. The observations were obtained with airmass values
between 1.1~and~3 and typical seeing values of 2~arcsec. In
Table~\ref{tab:observationDetail} we summarize the main
characteristics of our observations obtained at both sites. Combining
OLT and PTST data, the observations cover a total of 416~epochs.

\begin{table*}[ht!]
  \centering
  \caption[]{Summary of our observations carried out with OLT (top)
    and PTST (bottom) specifying epoch, exposure time ET, filter
    configuration F, number of data points NoP, airmass and transit
    coverage (TC); a description of the transit coding is detailed in
    the footnote of this table. Epochs are counted relative to the
    best-fitting transit. Transits observed simultaneously with both
    telescopes are marked in boldface.}
  \label{tab:observationDetail}
  \begin{tabular}{p{25mm} r p{4.5mm} p{4mm} p{4.5mm} l p{13mm}}
    \hline  
    \hline
    Date            &  Epoch  &  ET &  F & NoP & 
    Airmass & TC\\
                    &         &   (s)    &            &  &      &   \\ 
    OLT & & & & \\
    \hline
     2011 March 19    &  -364   &   202      & R$_1$      & 29         & 1.83$~\rightarrow~$1.40 &  - IBE -\\
     2011 March 26    &  -359   &   296      & R$_1$      & 28         & 1.45$~\rightarrow~$1.19 &  OIBE -\\ 
     2011 May 22      &  -319   &   173      & R$_1$      & 48         & 1.60$~\rightarrow~$1.18 &  OIBEO\\ 
     2011 May 29      &  -314   &    58      & R$_1$      & 72         & 1.19$~\rightarrow~$1.08 &  - - BEO\\ 
     2011 July 5      &  -288   &    82      & R$_1$      & 78         & 1.13$~\rightarrow~$1.03 &  - - BEO\\ 
     2011 July 15     &  -281   &   136      & R$_1$      & 86         & 1.21$~\rightarrow~$1.02 &  - - BEO\\ 
     2011 August 1st  &  -269   &    68      & R$_1$      & 108        & 1.12$~\rightarrow~$1.02 &  OIBEO\\ 
     2011 August 28   &  -250   &    80      & R$_1$      & 42         & 1.02$~\rightarrow~$1.04 &  - - BEO\\ 
     2011 October 1st &  -226   &    79      & R$_1$      & 136        & 1.12$~\rightarrow~$1.02 &  OIBEO\\ 
     2012 April 21    &   -83   &   141      & R$_1$      & 97         & 1.73$~\rightarrow~$1.18 &  OIBEO\\ 
     2012 April 24    &   -81   &    90      & R$_1$      & 40         & 1.80$~\rightarrow~$1.61 &  - - BEO\\ 
     2012 May 1st     &   -76   &    70      & R$_1$      & 177        & 1.71$~\rightarrow~$1.18 &  OIBEO\\  
     {\bf 2012 June 17}     &   -43   &    53      & R$_1$      & 61        & 1.39$~\rightarrow~$1.21 &- - - EO  \\ 
     2012 July 1st    &   -33   &    76      & R$_1$      & 112        & 1.12$~\rightarrow~$1.02 &  OIB - -\\
     {\bf 2012 August 17}   &     0   &    50      & R$_1$      & 425        & 1.12$~\rightarrow~$1.14 & OIBEO \\
     2012 September 13&    19   &    64      & R$_1$      & 124        & 1.05$~\rightarrow~$1.04 &  OI - - O\\
     2012 September 30&    31   &    59      & R$_1$      & 100        & 1.04$~\rightarrow~$1.14 &  - IBEO\\
     2012 October 30  &    52   &    52      & R$_1$      & 275        & 1.03$~\rightarrow~$1.16 &  OIBEO\\
    \hline
    PTST & & & & \\
    \hline
     2012 February 27 &  -121   &    90      & R$_2$      & 109        & 3.00$~\rightarrow~$1.47 &  OIBEO\\
     2012 May 11      &   -69   &    59      & R$_2$      & 183        & 2.84$~\rightarrow~$1.30 &  OIBEO\\
     2012 May 28      &   -57   &    90      & R$_2$      & 106        & 2.00$~\rightarrow~$1.18 &  OIBEO\\   
     {\bf 2012 June  17}    &   -43   &    40      & R$_2$      & 168        & 1.86$~\rightarrow~$1.34 & - - BEO \\  
     2012 July 4      &   -31   &    70      & R$_2$      & 80         & 1.50$~\rightarrow~$1.15 &  OIBEO\\  
     2012 July 14     &   -24   &    60      & R$_2$      & 86         & 1.42$~\rightarrow~$1.22 &  - - - EO\\
     2012 July 31     &   -12   &    60      & R$_2$      & 133        & 1.32$~\rightarrow~$1.11 &  OIBEO\\
     {\bf 2012 August 17}   &     0   &    40      & R$_2$      & 182        & 1.22$~\rightarrow~$1.23 & OIBEO \\
     \hline
  \end{tabular}
  \tablefoot{Observations were obtained in filter configurations R$_1$
    and R$_2$ denoting the Schuler Johnson-Cousins~R-band, and the
    Baader R-band, respectively. The letter code to specify the
      transit coverage during each observation is the following: O: out
      of transit, before ingress. I: ingress. B: flat bottom. E:
      egress. O: out of transit, after egress.}
\end{table*}

Calibration images such as bias and flat fields were obtained on each
observing night. We used the IRAF task {\it ccdproc} for bias
subtraction and flat-fielding on the individual data sets, followed by
the task {\it apphot} to carry out aperture photometry on all images
including individual photometric errors. We measured fluxes using
different apertures centered on the target star and six more stars
with a similar brightness as \qt, which are present in both field of
views. The apparent brightness of the reference star chosen to produce
the differential light curves is very close to \qt.  Multiband
photometry of the same field of view reveals no significant difference
between both stars as a function of photometric color, either
suggesting that they are of similar spectral type. This minimizes any
atmospheric extinction residuals on the differential light curves, and
in addition, the apparent proximity of the two stars ($\sim$2 arcmin)
minimizes systematic effects related to vignetting, comatic
aberration, or CCD temperature gradients, which all increase with
increasing distance from the telescope's optical axis (i.e., the
center of the chip).  We also checked the constancy of the reference
star against the other five comparison stars and chose as final
aperture the one that minimized the scatter in the resulting light
curves. The differential light curves were then produced by dividing
the flux of the target star by that of the reference star.

\begin{figure}[ht!]
  \centering
  \fbox{\includegraphics[width=.47\textwidth]{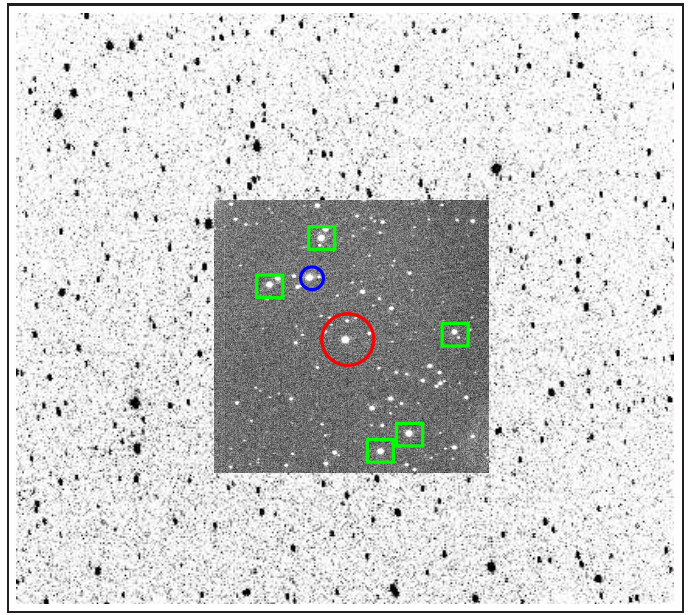}}
  \caption{\label{FoV} Oskar-L\"uhning telescope (dark background) and
    Planet Transit Search Telescope (light background) field of
    views. \qt \ is indicated with a large red circle centered on the OLT
    field of view, along with the comparison stars as green
    squares. The small blue circle upward and to right of \qt\ indicates the
    reference star used to produce the differential photometry.}
\end{figure}

Typical sky brightness values per binned pixel were of about 3500
counts for OLT, and 2000 for PTST.  Figure~\ref{FoV} shows the PTST
field of view (light background) superposed on the field of view of
OLT (dark background). \qt\ lies at the center of the field of view,
marked with a large red circle. The five comparison stars are
indicated with green squares and the reference star, indicated with a
small blue circle, is the one used to produce the differential light
curves.

After the reduction process, we fitted a straight line to the
out-of-transit data points to correct for any residual systematic
trend and to normalize the differential light curves. It is worth
mentioning that we calculated the timing offsets using both raw and
normalized data. We therefore confirm that the normalization process
does not produce any shifts in the mid-transits but only more accurate
planetary parameters. This might be because the amplitude of any
systematic trend present in our light curves was smaller than $\sim$2
mmag.

\section{Data analysis}
\label{sec:Analysis}

\subsection{Data preparation}

IRAF provides heliocentric corrections, therefore our time stamps are
given as heliocentric Julian dates (\hjdutc) and are converted into
barycentric Julian dates (\bjdtdb) using the web tool provided by
\citet{Eastman2010}\footnote{\url{http://astroutils.astronomy.ohio-state.edu/time/}}.

\subsection{Fit approach}

We fitted the transit data with the transit model developed by
\citet{MandelAgol2002}, making use of their \texttt{occultquad}
FORTRAN
routine\footnote{\url{http://www.astro.washington.edu/users/agol}}. From
the transit light curve, we can directly infer the following
parameters: the orbital period $P$, the mid-transit time $T_0$, the
radius ratio $p=R_p/R_s$, the semi-major axis (in stellar radii)
$a/R_s$, and the orbital inclination $i$.  For our fits we assumed a
quadratic limb-darkening law with fixed coefficients $u_1$ and $u_2$.

\subsection{Limb-darkening coefficients}

\citet{Alsubai2011} presented a spectroscopic characterization of \qt
\ based on comparing their observed spectrum with synthetic template
spectra and suggested that \qt \ has an effective temperature
$T_\mathrm{eff}$ of $4861\pm125$~K, a surface gravity $\log{g}$ of
$4.536\pm0.024$, and solar metallicity [Fe/H] of
$0.20\pm0.10$. Because our observations with OLT and PTST were
obtained using different (non-standard) filter sets, we decided to
calculate angle-resolved synthetic spectra from spherical atmosphere
models using PHOENIX (\citet{Peter1}, \citet{Peter2}) for a star with
effective temperature \mbox{$T_{\mathrm{eff}} = 4900$~K}, \mbox{[Fe/H]
  = 0.20} and \mbox{$\log{g} = 4.5$}, thereby closely matching the
spectroscopic parameters of \qta.  We then convolved each synthetic
spectrum with the OLT and PTST filter transmission functions and
integrated in the wavelength domain to compute intensities as a
function of $\mu = \cos{\theta}$, where $\theta$ is the angle between
the line of sight and the radius vector from the center of the star to
a reference position on the stellar surface. The thus derived
intensities were then fitted with a quadratic limb-darkening
prescription, viz.

\begin{equation*}
I(\mu)/I(1) = 1 - u_1(1 - \mu) - u_2(1 - \mu)^2\ ,
\end{equation*}

\noindent to obtain the $u_1$ and $u_2$ limb-darkening
coefficients. As an aside, we note that the best approach to obtain
limb-darkening coefficients would be to fit a more sophisticated
bi-parametric approximation to the stellar intensities
\citep{ClaretHauschildt2003} to the PHOENIX intensities, which
produces the smallest deviations in the generation of the
limb-darkening coefficients, which fits the stellar limb
better. However, to introduce such a limb-darkening law in the
production of primary transit light curves would be computationally
cumbersome. Moreover, \qt \ is a nearly grazing system, where changes
in limb darkening do not strongly affect the primary transit light
curves. We checked that the time that the planet spends in the
small-$\mu$ regime is very short. Furthermore, at a given time the
planet covers a broad range of $\mu$-values. Neglecting these points
in determining the limb-darkening coefficients will probably not
affect the subsequent parameter determination. To calculate the OLT
and PTST limb-darkening coefficients, we fitted a quadratic
limb-darkening law to PHOENIX intensities, neglecting the data points
between $\mu$ = 0 and $\mu$ = 0.1. Fig.~\ref{LD} shows the OLT and
PTST limb-darkening normalized functions, and Table~\ref{tab:LDcoefs}
the fitted limb-darkening coefficients.

\begin{figure}[ht!]
  \centering
  \includegraphics[width=.5\textwidth]{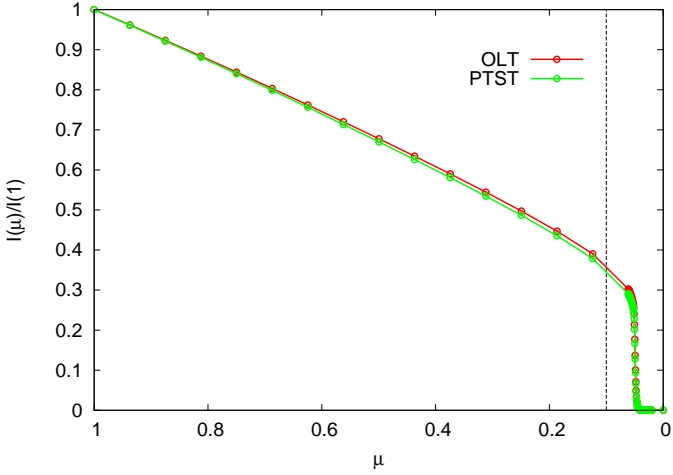}
  \caption{\label{LD} \qt \ normalized intensities considering the OLT and
    PTST filter transmission functions. Data points on the right of
    the vertical dashed line were not considered in the fitting
    procedure.}
\end{figure}

\begin{table}[t]
\caption[]{Best-fit limb-darkening coefficients (LDCs) for OLT
    and PTST, along with the 1 $\sigma$ errors.}
\label{tab:LDcoefs}
\centering
\begin{tabular}{lll}
\hline\hline
LDCs & OLT     & PTST \\
\hline
$u_1$ & 0.5860 $\pm$ 0.0053 & 0.6025 $\pm$ 0.0051 \\
$u_2$ & 0.1170 $\pm$ 0.0075 & 0.1140 $\pm$ 0.0073 \\
\hline
\end{tabular}
\end{table}

\subsection{Parameter errors}

To determine reliable errors for the fit parameters given the mutual
dependence of the model parameters, we explored the parameter space by
sampling from the posterior-probability distribution using a
Markov-chain Monte-Carlo (MCMC) approach.

The near-grazing transit geometry of \qt\ introduces a substantial
amount of correlation between the system parameters, which can easily
render the MCMC sampling process inefficient. To cope with these
difficulties we used a modification to the usually employed
Metropolis-Hastings sampling algorithm, which is able to adapt to the
strong correlation structure. This modified Metropolis-Hastings
algorithm is described in the seminal work by \citet{Haario2001} and
has become known as the adaptive Metropolis (AM) algorithm. It works
like the regular Metropolis-Hastings sampler, but relies on the idea
of allowing the proposal distribution to depend on the previous values
of the chain.

The regular Metropolis-Hastings algorithm produces chains based on a
proposal distribution that may depend on the current state of the
chain. The proposal distribution is often chosen to be a multivariate
normal distribution with fixed covariance. The main difference between
the regular Metropolis sampler and AM is that AM updates the
covariance matrix during the sampling, e.g. every 1000
iterations. Because the proposal distribution is thus tuned based on
the entire sampling run, AM chains in fact lose the Markov property.
Nonetheless, it can be shown that the algorithm retains the correct
ergodic properties under very general assumptions, \mbox{i.e.}, the
covariance matrix stabilizes during the sampling process and AM chains
properly simulate the target distribution \citep{Haario2001,
  Vihola2011}. Although adaptive MCMC algorithms are a recent
development, they have successfully been used by other authors in the
astronomical community, for instance by \cite{2009MNRAS.394.1936B,
  2010ApJ...718.1353I} and \cite{2011ApJ...742..123I}.

Our MCMC calculations make extensive use of routines of
\texttt{PyAstronomy}\footnote{\url{http://www.hs.uni-hamburg.de/DE/Ins/Per/Czesla/}\\ \url{PyA/PyA/index.html}},
a collection of Python routines providing a convenient interface for
fitting and sampling algorithms implemented in the PyMC
\citep{Patil2010} and SciPy \citep{Jones2001} packages. The AM sampler
is implemented in the PyMC package, which is publicly available for
download. We refer to the detailed online
documentation\footnote{\url{http://pymc-devs.github.io/pymc/}}.

We checked that AM does yield correct results for simulated data sets
with parameters close to \qt, and found that this approach showed fast
convergence and was efficient. To express our lack of more a priori
knowledge regarding the Qatar-1 system parameters, we assumed
uninformative uniform prior probability distributions for all
parameters, but we found that the parameters are well determined, so
that the actual choice of the prior is unimportant for our results.

\begin{figure*}\ContinuedFloat*
  \centering
  \includegraphics[width=0.9\textwidth]{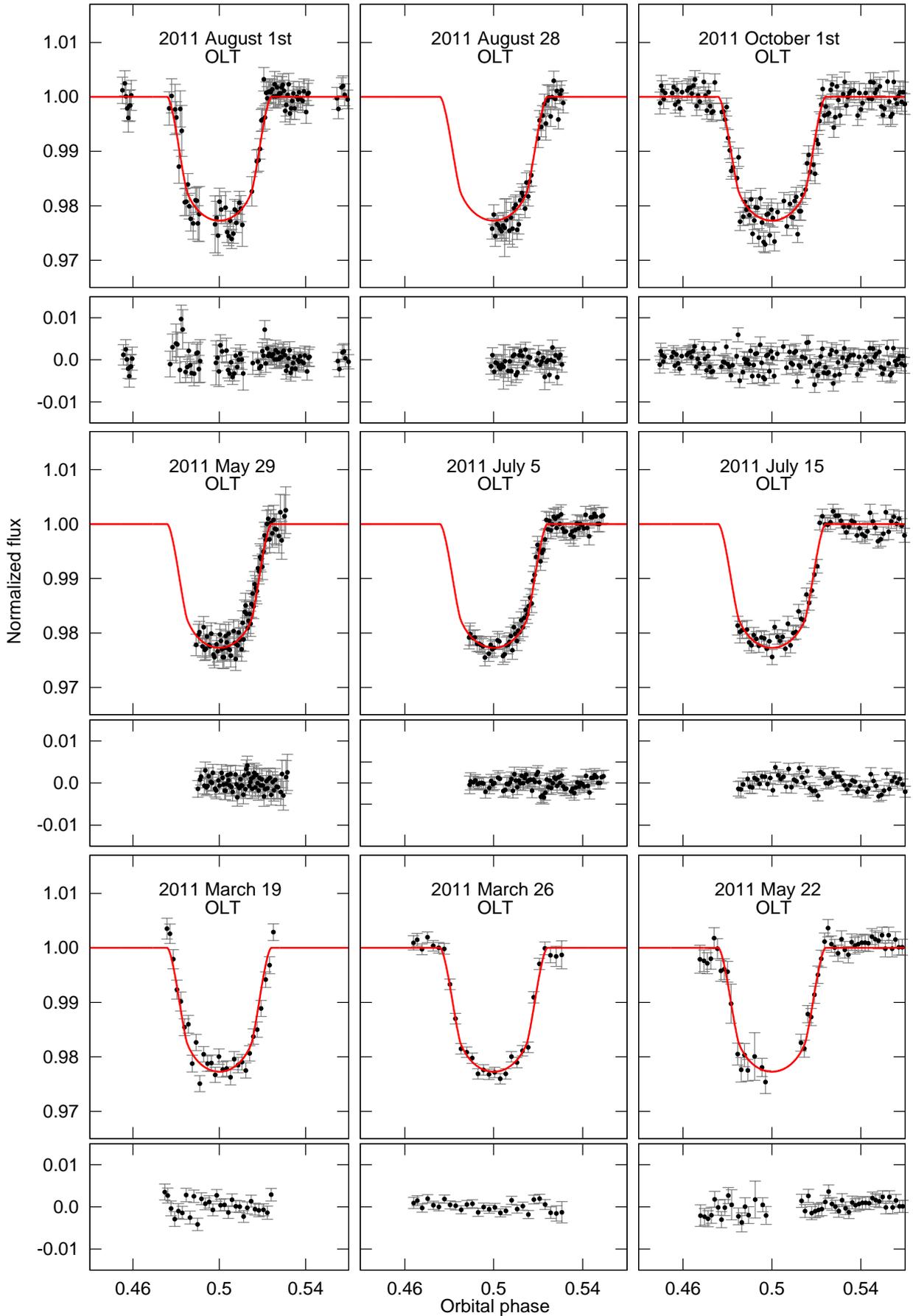}
  \caption{\label{fig:transits}OLT and PTST light curves (top
    panels). Superposed is the \citet{MandelAgol2002} primary transit
    light-curve model in a continuous line, considering the parameters
    obtained in Section 3.2. As indicated by the dates, the
    light curves evolve in time from bottom to top and from left to
    right. The residuals (bottom panels) have been calculated after
    subtracting the best transit fit parameters.}
\end{figure*}

\begin{figure*}\ContinuedFloat
  \centering
  \includegraphics[width=0.9\textwidth]{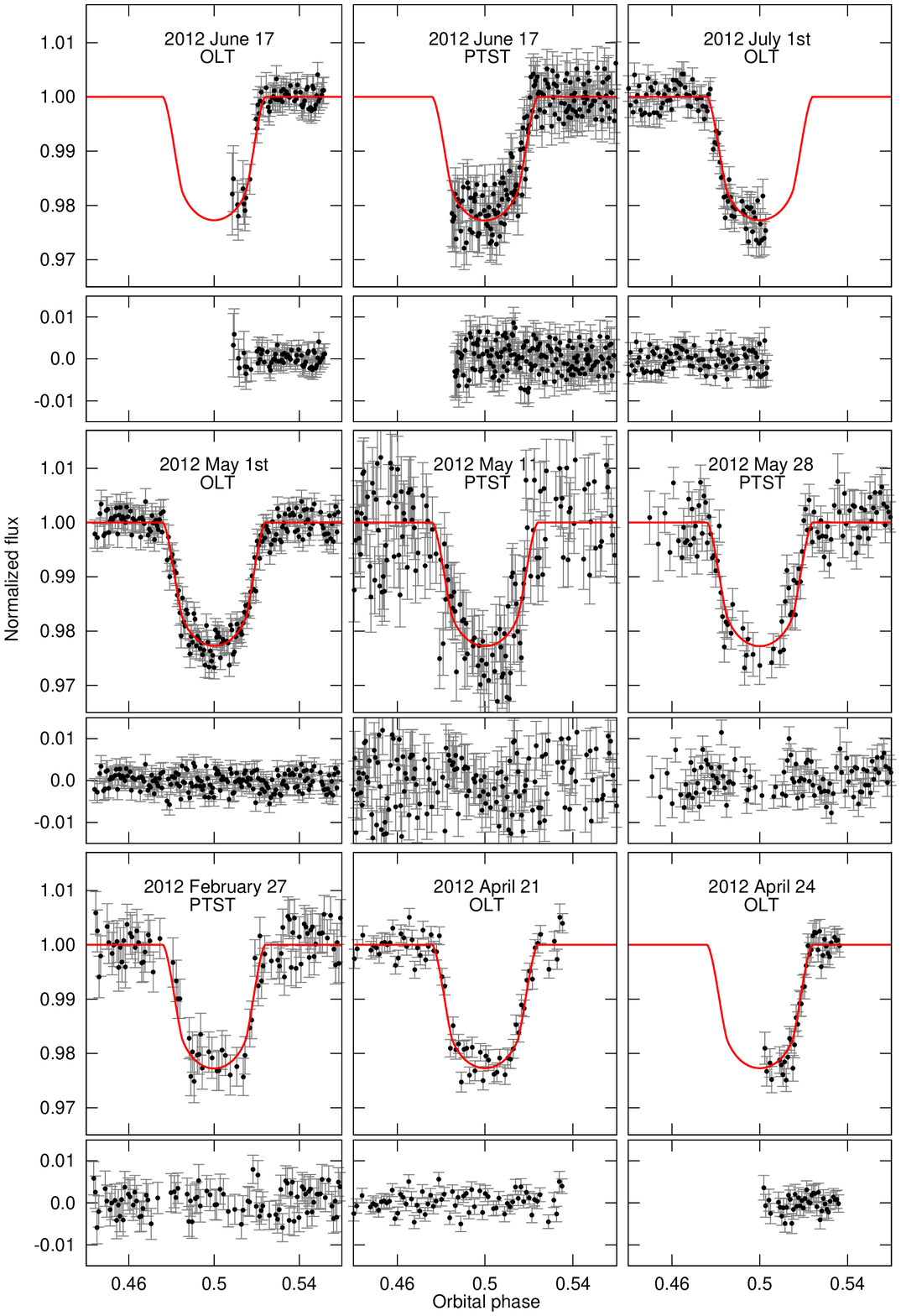}
  \caption{Same as Figure~\ref{fig:transits}}
\end{figure*}

\begin{figure*}\ContinuedFloat
  \centering
  \includegraphics[width=0.9\textwidth]{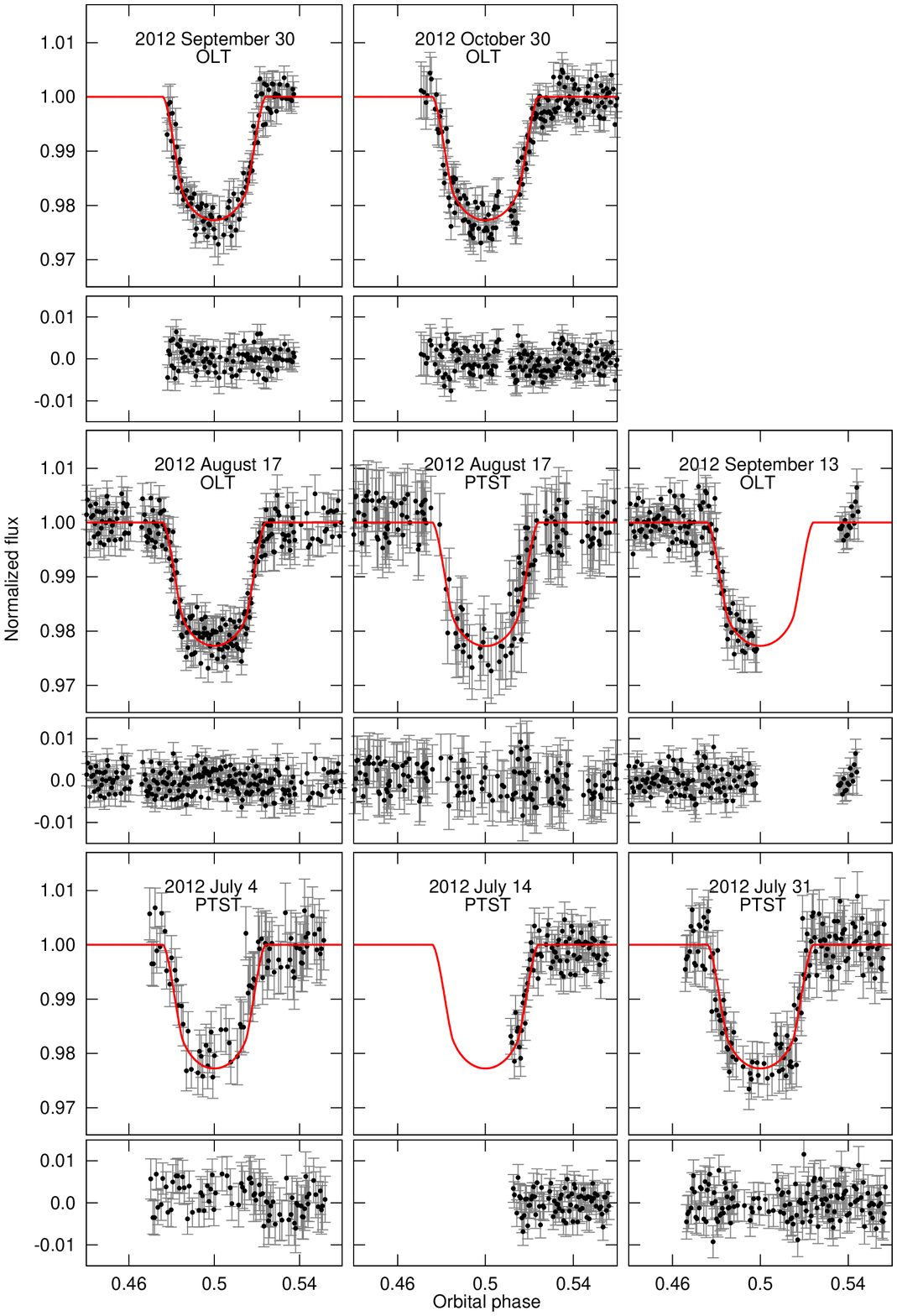}
  \caption{Same as Figure~\ref{fig:transits}}
\end{figure*}

In Figure~\ref{fig:transits} we show our 26 obtained light curves,
along with the residuals after removing the primary transit
feature. Our final relative photometry is available in its entirety in
machine-readable form in the electronic version of this paper. First
columns contain \bjdtdb, second columns normalized flux, and third
columns individual errors.

\subsection{Correlated noise}

\cite{Carter2009} (and references therein) studied how time-correlated
noise affects the estimation of the parameter of transiting systems.
To quantify whether and to what extent our light curves are affected
by red noise, we reproduced part of their analysis as follows: First,
we produced residuals from our final fits by subtracting the primary
transit model from each light curve. We individually divided each
light curve into M bins of equal duration. Since our data are not
always equally spaced, we calculated a mean value N of data points per
bin. If the data are not affected by red noise, they should follow the
expectation of independent random numbers,

\begin{equation*}
  \sigma_N = \sigma_1 N^{-1/2}[M/(M-1)]^{1/2}\ ,
\end{equation*}

\noindent where $\sigma_1$ is the sample variance of the unbinned data
and $\sigma_N$ is the sample variance (or RMS) of the binned data,
with the following expression:

\begin{equation*}
  \sigma_N = \sqrt{\frac{1}{M}\sum_{i = 1}^{M}(<\hat{\mu_i}> - \hat{\mu_i})^2}\ ,
\end{equation*}

\noindent where $\hat{\mu_i}$ is the mean value of the residuals per bin,
and $<\hat{\mu_i}>$ is the mean value of the means.

If correlated noise is present, then each value $\sigma_N$ will differ
by a factor $\beta_N$ from their expectation. The parameter $\beta$,
an estimation of the strength of correlated noise in the data, is
found by averaging $\beta_N$ over a range $\Delta n$ corresponding to
time scales that are judged to be most important. For data sets free
of correlated noise, we expect $\beta$ = 1. For primary transit
observations, $\Delta n$ is the duration of ingress or egress. For
\qt, the time between first and second contact (or equivalently, the
time between third and fourth contact) is $\sim$ 15 minutes.

\begin{figure}[ht!]
  \centering
  \includegraphics[width=.49\textwidth]{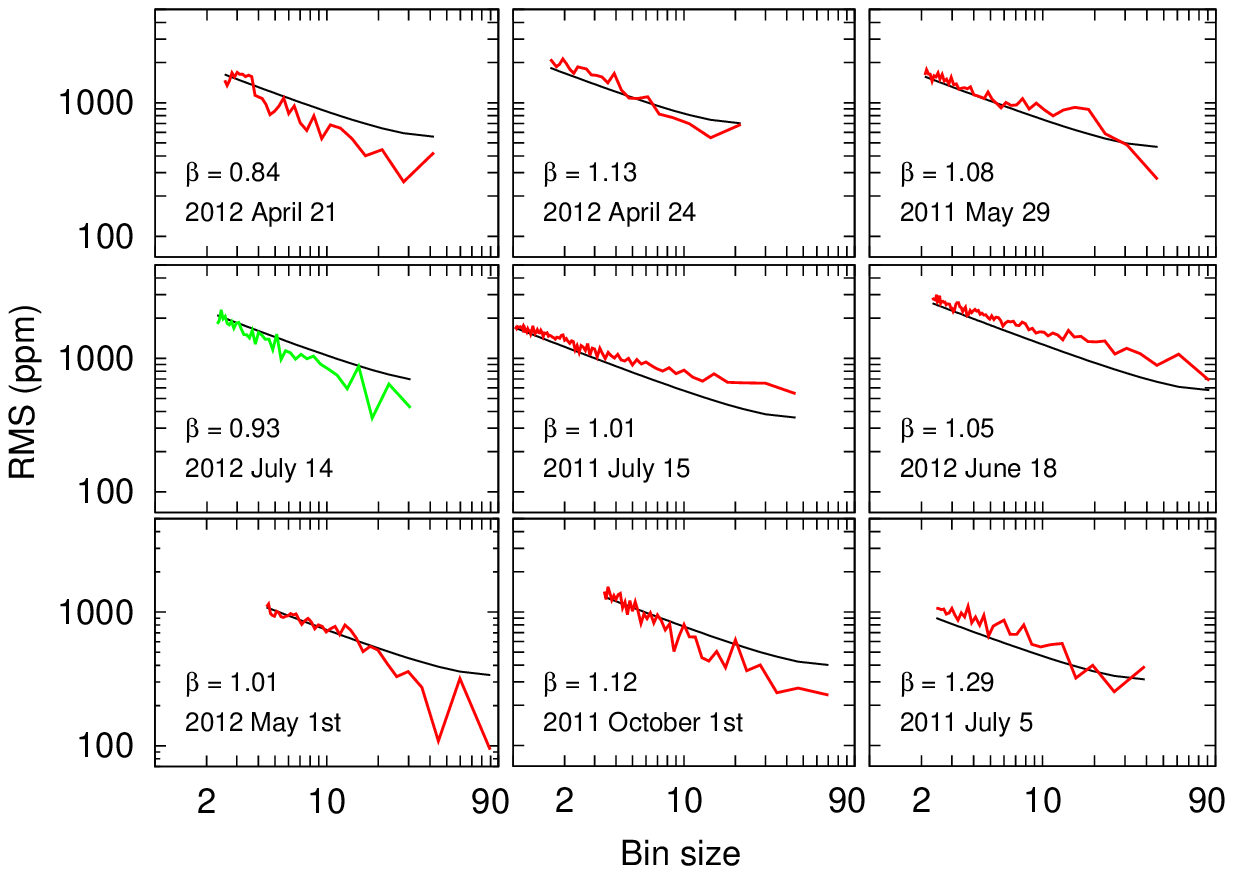}
  \caption{\label{corrNoise} \qt \ RMS in parts per Million (ppm) of
    the time-binned residuals as a function of bin size in logarithmic
    scale. Red and green lines correspond to OLT and PTST data
    respectively, and black lines show the expected behavior under the
    presence of uncorrelated noise.}
\end{figure}

In Figure~\ref{corrNoise} we show the results of our correlated noise
analysis for nine of the longest light curves. Black lines represent
the expected behavior in the absence of red noise, and red and green
lines represent the variance of the binned data for OLT and PTST,
respectively, as a function of bin size. As expected, the larger the
bin size, the smaller the RMS. For each light curve we calculated
$\beta$ considering $\Delta n$ = 15 minutes. For the OLT and PTST
primary transits, $\beta$ lies between $\beta$ = 0.78 and $\beta$ =
1.33, with $< \beta >$ = 1.038. Thus, there is no evidence for
significant correlated noise in our light curves.

Finally, \cite{Pont2006} suggested to enlarge individual photometric
errors by a factor $\beta$ to account for systematic effects on the
light curves. This would increase the parameter errors without
changing the parameter estimates. Since our light curves do not
present any strong evidence of correlated noise, we did not modify the
individually derived errors.

\subsection{Effects of exposure times and transit coverage}

The accuracy in determining of the mid-transit time is affected by the
number of data points during transit and the number of off-transit
data points, even more so when a normalization is involved.

\begin{figure}
  \centering
  \includegraphics[width=.49\textwidth]{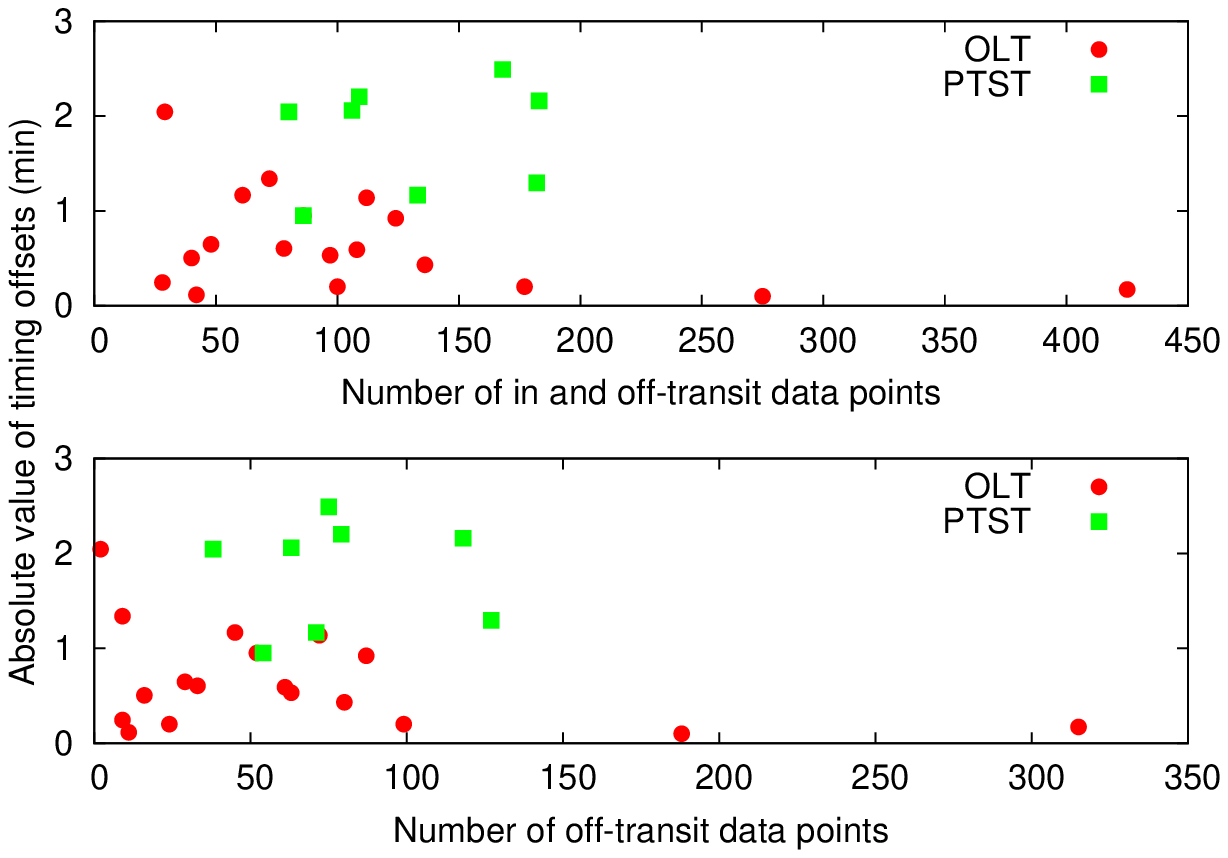}
  \caption{\label{fig:transitDur} Timing-residual magnitudes in
    minutes as a function of in and off-transit data point number
    (top) and off-transit data point number only (bottom). The
    outermost point close to zero is the primary transit, which was selected to be
    the $0^{th}$ epoch. Mid-transit errors are not plotted to avoid
    visual contamination.}
\end{figure}

To study whether the mid-transits are affected by the transit
observation duration, we computed the timing residual magnitudes as a
function of the number of data points per transit. If any systematic
effect dominates the light curves, a larger timing offset for those
transits that are sampled the least is expected. We show our results
in Figure~\ref{fig:transitDur}. Since the light curves are normalized
and the mid-transits might be sensitive to normalizations, we also
computed the timing-residual magnitudes as a function of the number of
off-transit data points. We used the Pearson correlation coefficient
$r$ to quantify the correlation of the mid-transit offsets. In both
cases this was \mbox{$r \sim$ -0.11} for all data points, and \mbox{$r
  \sim$ 0.05}, which rules out the 0$^{th}$ epoch, which was expected
to be zero by construction. We found no significant correlation.

\begin{figure}
  \centering
  \includegraphics[width=.49\textwidth]{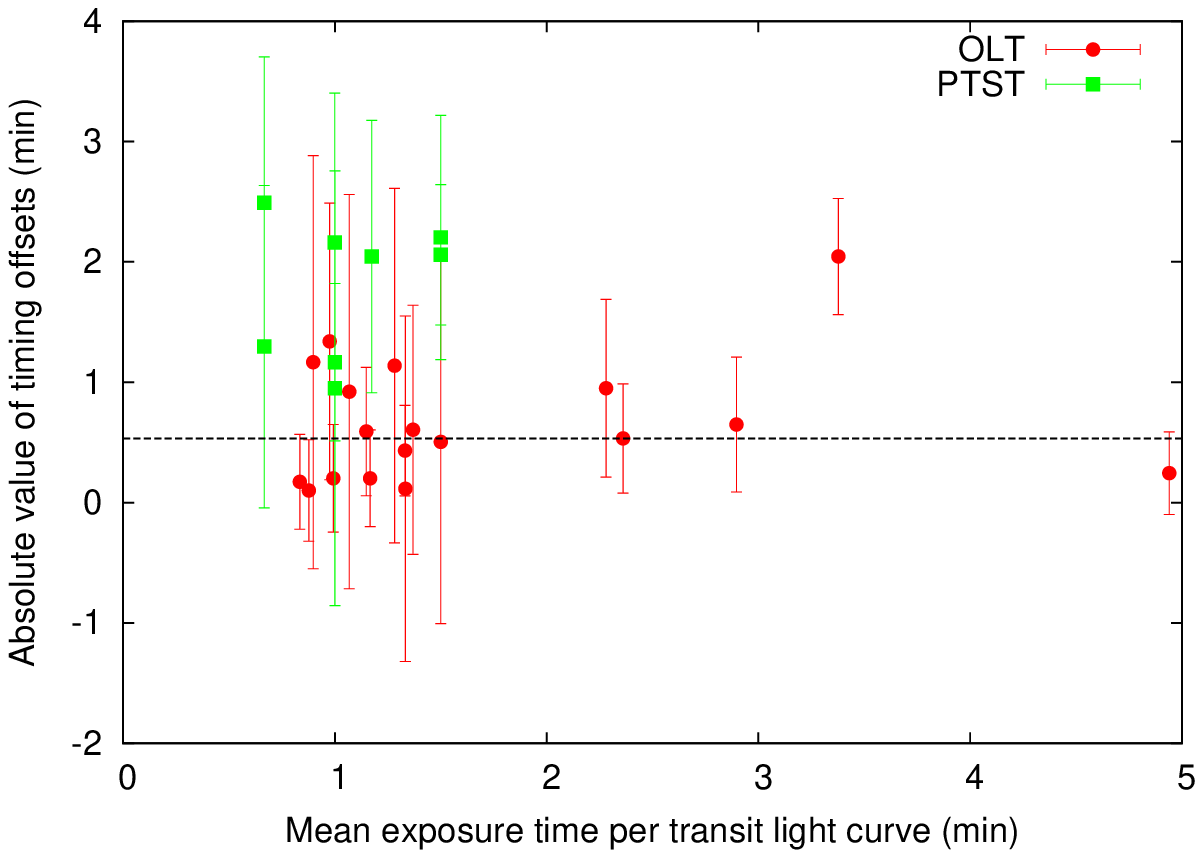}
  \caption{\label{fig:cadence} Timing-residual magnitudes in minutes
    for OLT (red circles) and PTST (green squares), as a function of
    mean exposure times. The horizontal dashed line indicates half of
    the mean exposure time.}
\end{figure}

\cite{Kipping2010} studied the effects of finite integration times on
the determination of the orbital parameters. He showed that the time
difference between the mid-transit moment and the nearest light curve
data point might cause a shift in the TTV signal, which is expected to
be one half of the rate sampling. To test how significant this effect
is over our light curves, we calculated the mean exposure times per
observing night, which was about 80 seconds. Figure~\ref{fig:cadence}
shows the mean exposure times per night, versus the magnitude of the
timing residuals. Most of the data points lying around the half-mean
exposure time ($\sim$ 35 seconds) were identified to have the most
accurate mid-transit timings, with exposure times of about one minute,
which makes it unlikely that they are affected by a sampling
effect. The Pearson correlation coefficient of \mbox{$r$ = -0.15}
again reveals no significant correlation.

\subsection{Results}

In general terms, the parameters $P$ and $T_0$ are usually correlated
with each other, but are uncorrelated with the remaining transit
parameters. Therefore, these parameters can be determined quite
accurately without interference of the remaining ones. We used a
Nelder-Mead simplex to approach the best-fit solution, which is
provided as the starting values of the MCMC sampler. The AM algorithm
then samples from the posterior distribution for the parameters to
obtain error estimates. After $10^8$ iterations we discarded a
suitable burn-in (typically $10^6$ samples) and determined the
combination of parameters resulting in the lowest deviance. We
consider the lowest deviance as our global best-fit solution. The
errors were derived from the 68\,\% highest probability density or
credibility intervals (1 $\sigma$). Our results are summarized in
Table~\ref{tab:results}. None of our fit parameters are consistent
with those reported by \citet{Alsubai2011}; a possible explanation for
this inconsistency might be that these authors determined the system
parameters using only four light curves, two of which were incomplete,
obtained in less than two weeks. During the revision of this paper,
\cite{Covino2013} presented high-precision radial velocity
measurements from which the Rossiter-McLauglin effect was
observed. The authors also obtained five new photometric transit light
curves, from which the orbital parameters of the system were
improved. With respect to the orbital period, one of the most
important parameters for determining TTVs, our best-fitted orbital
period seems to be consistent within errors with the one found by
\cite{Covino2013}.

\begin{table}[t]
\caption[]{Best-fit (lowest deviance) parameters for our 26 transits
  of \qt \ after $10^8$ MCMC samples. LDCs are fixed at the
  values reported in Table~\ref{tab:LDcoefs}.}
\label{tab:results}
\centering
\begin{tabular}{lll}
\hline\hline
Parameter & Value (1$\sigma$ errors)      & \citet{Alsubai2011} \\
\hline
$i$ ($^{\circ}$) & 84.52 $\pm$ 0.24          & 83.47$^{+0.40}_{-0.36}  $    \\
$p$ ($R_p/R_s$) & 0.1435 $\pm$ 0.0008         & 0.1454 $\pm$ 0.0015   \\
$a$ ($R_s$)    & 6.42 $\pm$ 0.10             & 6.1005$^{+0.067}_{-0.065}$    \\
$P$ (days)     & 1.4200246 $\pm$ 0.0000004   &  1.420033 $\pm$ 0.000016   \\
\hline
\end{tabular}
\end{table}

To obtain individual mid-transit times $T_{mid,i}$, we considered the
best-fit values of $p$, $a$, $i$, $u_1$, $u_2$, and $P$. We specified
Gaussian priors on these parameters and refitted each one of the 26
individual transit light curves for the mid-transit time
$T_{mid,i}$. Our results are listed in Tab.~\ref{tab:OminusC} together
with the derived 1 $\sigma$ errors on these transit times.  To compute
the timing deviations compared with a constant period we fitted the
observed mid-transit times $T_{o,i}$ to the expression

\begin{equation*}
  T_{o,i} = P \cdot E_i + T_o\ ,
\end{equation*}

\noindent finding the ephemeris 

$$P = 1.4200246 \pm 0.0000004~ \mbox{days}$$
$$T_0 = 2456157.42204 \pm 0.0001~ \mbox{\bjdtdb}$$

\noindent as best-fitting values. All errors are obtained from the
68.27\% confidence level of the marginalized posterior distribution
for the parameters. With these parameters we computed the OC-values
also listed in Tab.~\ref{tab:OminusC}. We used the available
simultaneous observations separated two months from each other as a
diagnostics of our fitting procedure. Both mid-transits (epochs -43
and 0) are consistent with each other within the errors.

\section{Transit timing variation analysis}
\label{sec:TTV}

\subsection{OC diagram}

In the absence of any timing variations we expect no significant
deviations of the derived OC-values from zero. Testing the null
hypothesis OC = 0 with a $\chi ^2$-test, we found
\mbox{${\chi}^2_{red}$ = 2.56} with 24 degrees of freedom.  This high
value led us to reject the null hypothesis that OC vanishes.

\begin{figure}[ht!]
  \centering
  \includegraphics[width=0.49\textwidth]{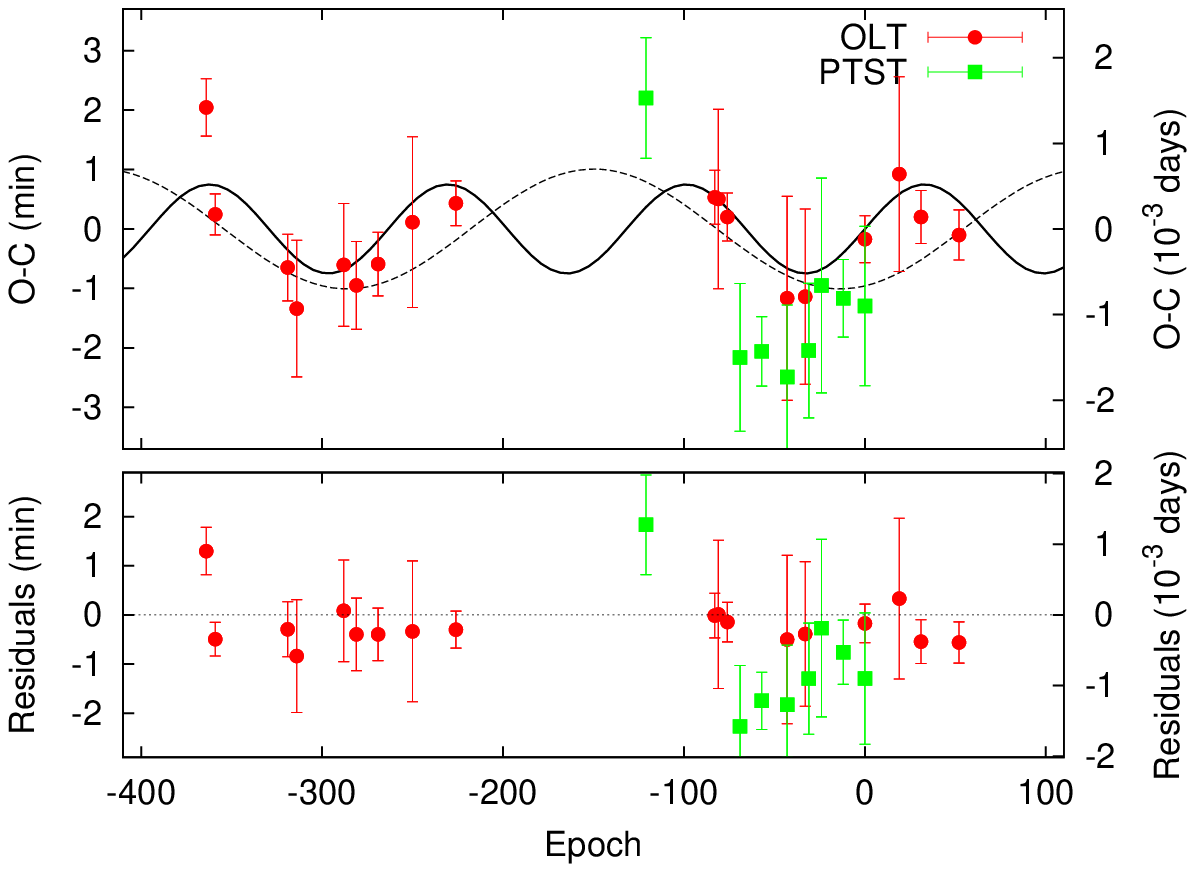}
  \caption{\label{fig:OminusC}OC diagram for \qt\ in minutes
      (left axis) and days (right axis) considering the OLT (red circles)
    and PTST (green squares) data points, along with the timing
    residuals. Our initial best-fitting model is overplotted with
    a continuous black line, and the second one with a dashed black line.}
\end{figure}

We then applied the Lomb-Scargle periodogram (\cite{Lomb},
\cite{Scargle}, \cite{LombScargle}) to search for any significant
periodicity contained in the OC diagram.  Figure~\ref{fig:OminusC}
shows the data points used to perform the periodogram, along with a
list in Table~\ref{tab:OminusC}.  The resulting periodogram reveals a
first peak at \mbox{$\nu_{TTV,1}$ = 0.00759 $\pm$ 0.00075 cycl
  P$^{-1}$} (corresponding to a period of 187 $\pm$ 17 days) and a
second one at about half the frequency \mbox{$\nu_{TTV,2}$ = 0.00367
  $\pm$ 0.00059 cycl P$^{-1}$} (corresponding to 386 $\pm$ 54
days). For both periodic signals, new ephemeris were refitted using
the linear trend and a sinusoidal variation in the form

\begin{equation*}
  T_{o,i} = P \cdot E_i + T_o + A_{TTV}\sin(2 \pi \nu_{TTV}[E_i - E_{TTV}]).
\end{equation*}

\begin{table}[ht!]
\caption[]{\label{tab:OminusC}OLT (top) and PTST (bottom) fitted
  mid-transits with 1 $\sigma$ errors and the OC data points.}
\centering
\begin{tabular}{r r r}
\hline\hline
Epoch & $T_0$                                 & O-C \\
      & \bjdtdb - 2455500              & (days) \\
OLT   &                                       &   \\
\hline 
-364  & 140.53449  $\pm$ 0.00033 & 0.00142   \\
-359  & 147.63337  $\pm$ 0.00023 & 0.00017  \\
-319  & 204.43373  $\pm$ 0.00038 & -0.00045  \\
-314  & 211.53337  $\pm$ 0.00079 & -0.00093  \\
-288  & 248.45453  $\pm$ 0.00071 & -0.00042   \\
-281  & 258.39446  $\pm$ 0.00051 & -0.00066  \\
-269  & 275.43500  $\pm$ 0.00037 & -0.00041  \\
-250  & 302.41596  $\pm$ 0.00099 & 0.00007   \\
-226  & 336.49677  $\pm$ 0.00026 & 0.00030   \\
-83   & 539.56037  $\pm$ 0.00031 & 0.00037   \\
-81   & 542.40039  $\pm$ 0.00104 & 0.00035   \\
-76   & 549.50031  $\pm$ 0.00027 & 0.00014   \\
-43   & 596.36017  $\pm$ 0.00119 & -0.00081  \\
-33   & 610.56044  $\pm$ 0.00102 & -0.00079  \\
0     & 657.42192  $\pm$ 0.00027 & -0.00012  \\
19    & 684.40315  $\pm$ 0.00113 & 0.00064   \\
31    & 701.44294  $\pm$ 0.00031 & 0.00014  \\
52    & 731.26325  $\pm$ 0.00029 & -0.00007  \\
\hline
PTST  &  &   \\
\hline
-121  & 485.60059  $\pm$ 0.00070 & 0.00153  \\
-69   & 559.43884  $\pm$ 0.00086 & -0.00150  \\
-57   & 576.47921  $\pm$ 0.00040 & -0.00143  \\
-43   & 596.35925  $\pm$ 0.00084 & -0.00173  \\
-31   & 613.39986  $\pm$ 0.00078 & -0.00142  \\
-24   & 623.34079  $\pm$ 0.00125 & -0.00066  \\
-12   & 640.38093  $\pm$ 0.00045 & -0.00081  \\
0     & 657.42114  $\pm$ 0.00093 & -0.00090  \\
\hline
\end{tabular}
\end{table}

For the most significant frequency, the fitted amplitude and phase are
\mbox{$A_{TTV,1}$ = 0.00052 $\pm$ 0.0002 days} (equivalently,
0.75$\pm$ 0.28 minutes) and \mbox{$\phi_{TTV,1}$ = 0.04 $\pm$
  0.05}. We then recalculated \mbox{$\chi^2_{red}$ = 1.85} for the
first frequency. Using an F-test to check the significance of the fit
improvement we obtained a p-value of 0.02, indicating that the
sinusoidal trend does indeed provide a better description than the
constant at 2.8$\sigma$ level. We can also compare the models using
the Bayesian information criterion, \mbox{BIC = $\chi^2$ + k ln N},
which penalizes the number k of model parameters given N = 26 data
points. We obtained BICs of 68.1 and 57.2 for the constant and
sinusoidal trend, respectively. Since the BIC is no more than a
criterion for model selection among a set of models, the method one
more time favors the sinusoidal trend.

We finally estimated the false-alarm probability (FAP) of the TTV
signal, using a bootstrap resampling method by randomly permuting the
mid-transit values ($5\times10^5$ times) along with their individual
errors, fixing the observing epochs and calculating the Lomb-Scargle
periodogram afterward. We estimated the FAP as the frequency with
which the highest power in the scrambled periodogram exceeds the
maximal power in the original periodogram.  In this fashion we
estimated an FAP of 0.05\% for our observed TTV signal, consistent
with our previous estimates.

To further check whether the addition of the sinusoidal variation
provides an explanation for the timing offsets, we made use of the
mid-transits of \qt \ available in the Exoplanet Transit
Database\footnote{\url{http://var2.astro.cz/ETD/}} (ETD) \citep{ETD}
to reinforce our results.  The ETD data are very heterogeneous, we
therefore converted the best 26 mid-transits from \hjdutc \ to
\bjdtdb. Each ETD primary transits has a data quality indicator
ranging from 1 (small scatter and good time-sampling) to 4 (the
opposite). For our analysis we only selected the transits with quality
flags 1 (5 primary transits) and 2 (21 primary transits).  These
transits span a total of $\sim$ 500 epochs. We produced the OC diagram
(Table~\ref{tab:ETD_OC}) using the following fitted ephemeris:

  $$P = 1.4200223 \pm 0.0000004~ \mbox{days}$$ 
  $$T_0 = 2455518.410961 \pm 0.0002~\mbox{\bjdtdb}.$$

\begin{table}[ht!]
\caption[]{\label{tab:ETD_OC}ETD-fitted 
  instant of minima with 1 $\sigma$ errors and the OC data points.}
\centering
\begin{tabular}{r r r}
\hline\hline
Epoch & $T_0$                                 & O-C \\
      & \bjdtdb - 2455500                  & (days) \\
ETD   & & \\
\hline 

0    &   18.41096 $\pm$ 0.00020  &  0\\
93   &  150.47263 $\pm$ 0.00058  &  -0.00040\\
119  &  187.39490 $\pm$ 0.00063  &  0.00128\\
155  &  238.51887 $\pm$ 0.00062  &  0.00445\\
158  &  242.77532 $\pm$ 0.00088  &  0.00083\\
189  &  286.79573 $\pm$ 0.00059  &  0.00055\\
194  &  293.89601 $\pm$ 0.00041  &  0.00072\\
196  &  296.73493 $\pm$ 0.00039  &  -0.00040\\
219  &  329.39461 $\pm$ 0.00049  &  -0.00123\\
231  &  346.43651 $\pm$ 0.00054  &  0.00039\\
238  &  356.37502 $\pm$ 0.00077  &  -0.00124\\
246  &  367.73872 $\pm$ 0.00054  &  0.00227\\
264  &  393.29753 $\pm$ 0.00051  &  0.00068\\
265  &  394.71796 $\pm$ 0.00049  &  0.00109\\
329  &  485.59892 $\pm$ 0.00067  &  0.00062\\
341  &  502.63734 $\pm$ 0.00044  &  -0.00122\\
401  &  587.84221 $\pm$ 0.00053  &  0.00230\\
412  &  603.46031 $\pm$ 0.00058  &  0.00016\\
419  &  613.40040 $\pm$ 0.00038  &  0.00009\\
434  &  634.70330 $\pm$ 0.00038  &  0.00266\\
443  &  647.47797 $\pm$ 0.00041  &  -0.00286\\
446  &  651.74233 $\pm$ 0.00046  &  0.00142\\
450  &  657.41961 $\pm$ 0.00049  &  -0.0013\\
455  &  664.52112 $\pm$ 0.00043  &  0.00001\\
495  &  721.32109 $\pm$ 0.00073  &  -0.00090\\
510  &  742.62146 $\pm$ 0.00048  &  -0.0008\\
\hline
\end{tabular}
\end{table}

\begin{figure}[ht!]
  \centering
  \includegraphics[width=0.5\textwidth]{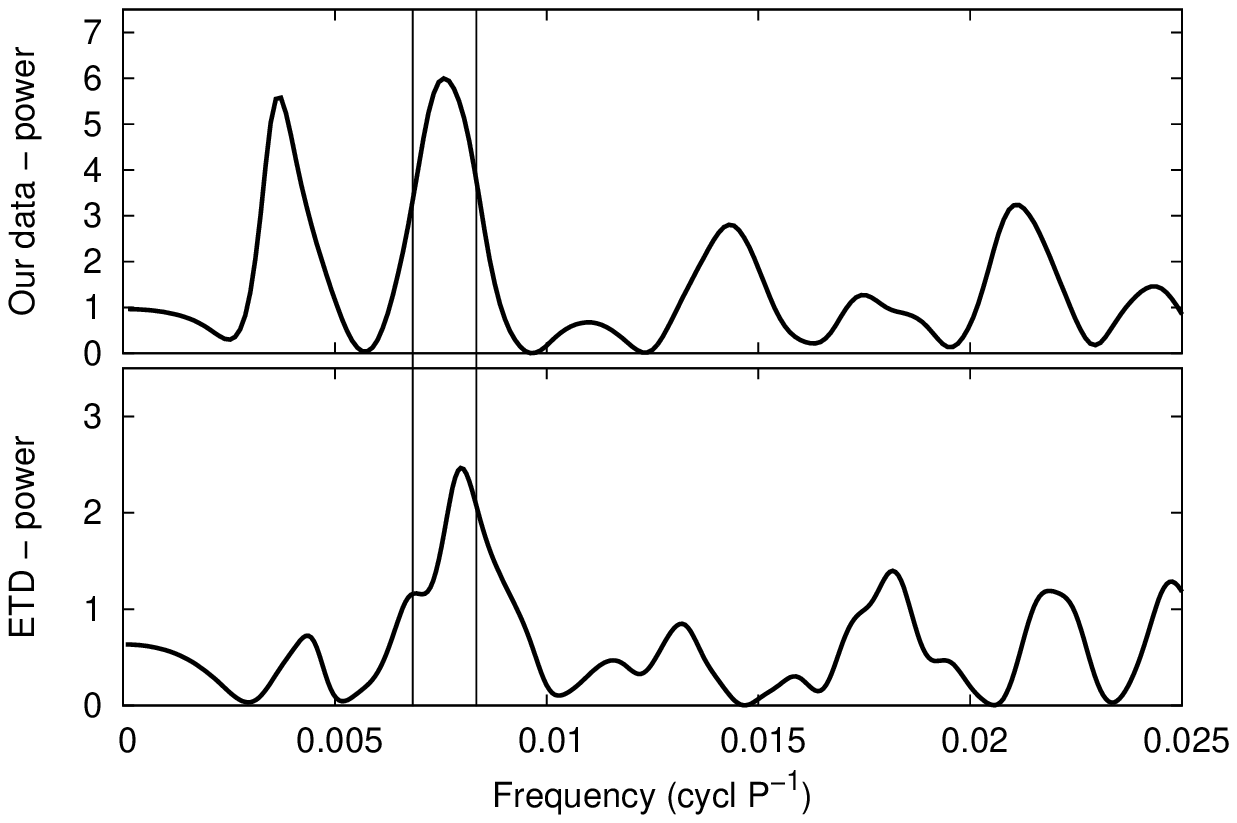}
  \caption{\label{fig:Periodogram}\cite{LombScargle} periodograms
    generated from the timing residuals of \qt, showing a peak at
    \mbox{$\nu_{TTV,1}$ = 0.00759 $\pm$ 0.00075 cycl P$^{-1}$} (our data,
    top panel) and \mbox{$\nu_{TTV,ETD}$ = 0.0078 $\pm$ 0.0011 cycl
    P$^{-1}$} (ETD data, bottom panel). Vertical lines indicate
    the 1$\sigma$ error on $\nu_{TTV,1}$. The parameters derived from our
    periodogram are maximum power = 6.1, and FAP for the maximum-power peak
    = 0.19\%.}
\end{figure}

The primary transit selected for the 0$^{th}$ epoch is the most
accurate one of all available light curves. We produced a periodogram
using the resulting timing offsets and found one peak at
\mbox{$\nu_{TTV,ETD}$ = 0.00784 $\pm$ 0.0011 cycl P$^{-1}$}
(corresponding to a period of 181 $\pm$ 22 days) with an amplitude of
\mbox{$A_{TTV,ETD}$ = 0.0009 $\pm$ 0.0005 days} (equivalently,
\mbox{1.29 $\pm$ 0.72 minutes}) and a phase of \mbox{$\phi_{TTV,ETD}$
  = 0.02 $\pm$ 0.03}. Within the errors, the frequency, amplitude, and
phase are consistent with the ones found using OLT and PTST
data. Figure~\ref{fig:Periodogram} shows two periodograms, the one on
top produced using our data, and the one on bottom produced with ETD
data.

\subsection{Error analysis}

To test the reliability of our error estimates for each mid-transit
time and to check the influence of the error magnitude on the
estimated peak frequencies, we iteratively constructed new
Lomb-Scargle periodograms randomly increasing the individual
mid-transit errors mostly by a factor of two.

At each iteration, we first added the absolute value of a random
number that was drawn from a normal distribution ($\mu$ = 0, $\sigma$
= 0.0004 days) to each mid-transit error, and calculated the leading
frequency afterward. After $5\times 10^5$ of such iterations, we found
that the only effect on the error increment is the permutation of the
leading frequency from $\nu_{TTV,1}$ to $\nu_{TTV,2}$. This
permutation occurred only $\sim$9\% of the times.

\begin{figure}[ht!]
  \centering
  \includegraphics[width=.5\textwidth]{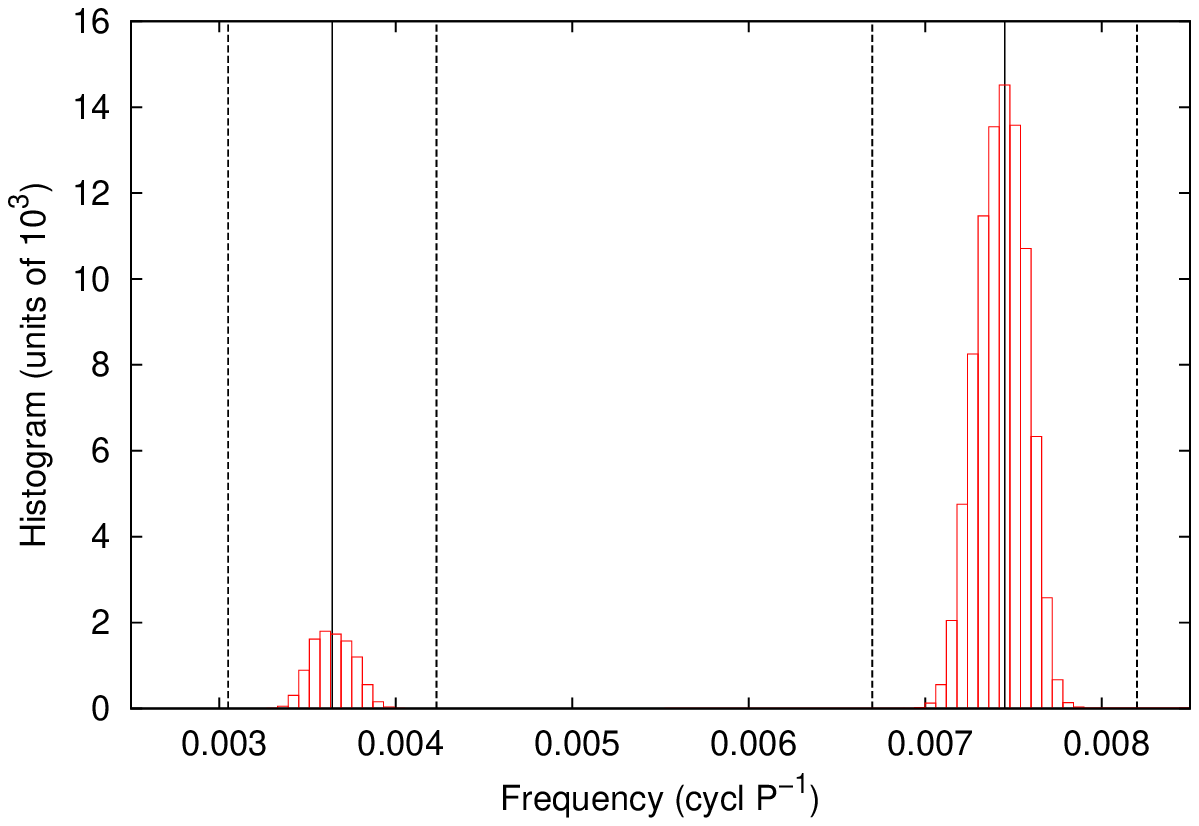}
  \caption{\label{fig:histFreq} Histogram of the leading frequencies
    obtained after incrementing the individual mid-transit errors, in
    units of $10^{3}$. Vertical continuous lines indicate our best two
    leading frequencies $\nu_{TTV,1}$ and $\nu_{TTV,2}$, along with
    1$\sigma$ errors (dashed vertical lines).}
\end{figure}

In Figure~\ref{fig:histFreq} we show the resulting histogram of the
calculated leading frequencies. They all fall inside the 1$\sigma$
errors of $\nu_{TTV,1}$ and $\nu_{TTV,2}$, estimated from our original
fits. Thus, the derived mid-transit errors do not significantly affect
the outcome in terms of dominating frequencies.

\section{Interpretation of the observed transit timing variations}
\label{sec:dyn}

For the purposes of this section we assumed that the reported TTVs are
real and explored possible physical scenarios that would explain the
observed variations. We considered both the $\sim$190 and $\sim$380
day periods, although the ETD data set is consistent only with the
former period. As is well known, a suitably placed third body in the
Qatar-1 system can lead to perturbations in the orbit of Qatar-1b,
which manifest themselves as TTVs.  Since there are no complete
analytical solutions to the three-body problem, we made use of a
numerical integration scheme. Because a planetary orbit is determined
by a large number of parameters (i.e., the eccentricity, the longitude
of the nodes, the orbital inclination, among many others), it is a
real challenge to estimate true orbital parameters by fitting TTV
models to ground-based mid-transit shifts only. We therefore started
with the simplest approach and assumed coplanar orbits using the
inclination derived from the light curve fitting, along with the
best-fit orbital period, and the transiting planet mass and
eccentricity obtained by \cite{Alsubai2011} as fixed. Given trial
mass, orbital period, eccentricity, longitude of periastron and
ascending node, and the time of periastron passage, the n-body code
calculates the time of occurrence of the primary transits taking into
account the interaction between the planets.  We specifically
considered two scenarios, a putative low-mass planet in a resonant
orbit relative close to Qatar-1b, and a putative high-mass planet or
brown dwarf in an 190-day orbit around Qatar-1.

\subsection{Weak perturber in resonance with \qtb}

Two planetary bodies in orbital resonance with each other will
experience long-term changes in their orbital parameters. On one hand,
since only a relative short stretch of data (less than two years) is
available, expecting a unique solution is too ambitious. On the other
hand, most of the perturber's dynamical setups will translate into
negligible TTVs.

\begin{figure}
  \centering
  \includegraphics[width=.5\textwidth]{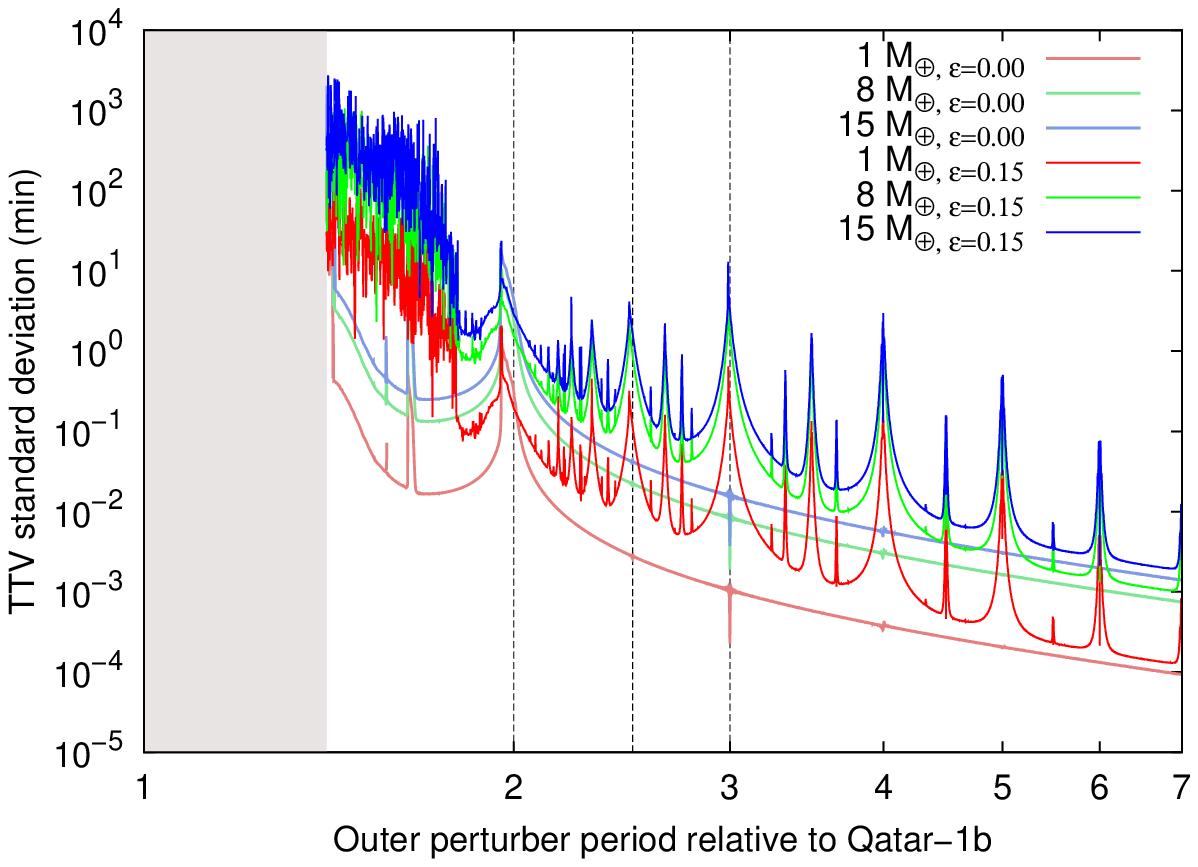}
  \caption{\label{fig:scatter} Simulated TTV standard deviation in
    minutes of \qtb \ as a function of the outer perturber period, for
    two values of eccentricity and three values of perturber masses. The
    gray area on the left correspond to perturber orbits that would
    make the \qtb \ orbit unstable. The mean motion resonances 2:1, 3:1,
    and 5:2 are indicated with vertical dashed lines.}
\end{figure}

Taking into account these difficulties, we calculated the modeled TTV
scatter for a two-planetary system, to study in advance which would be
the parameter space giving rise to TTVs similar to the scatter of our
data. We first considered both bodies in circular orbits and then the
perturber in an eccentric orbit. The mass of the perturber was
systematically changed between 1, 8, and 15 $M_{\oplus}$.
Figure~\ref{fig:scatter} shows our results. The vertical dashed lines
indicate the 2:1, 3:1, and 5:2 resonances. In these regions,
perturbers of the order of several Earth masses would produce TTVs
similar to the observed ones.

\begin{figure}[ht!]
  \centering
  \includegraphics[width=.5\textwidth]{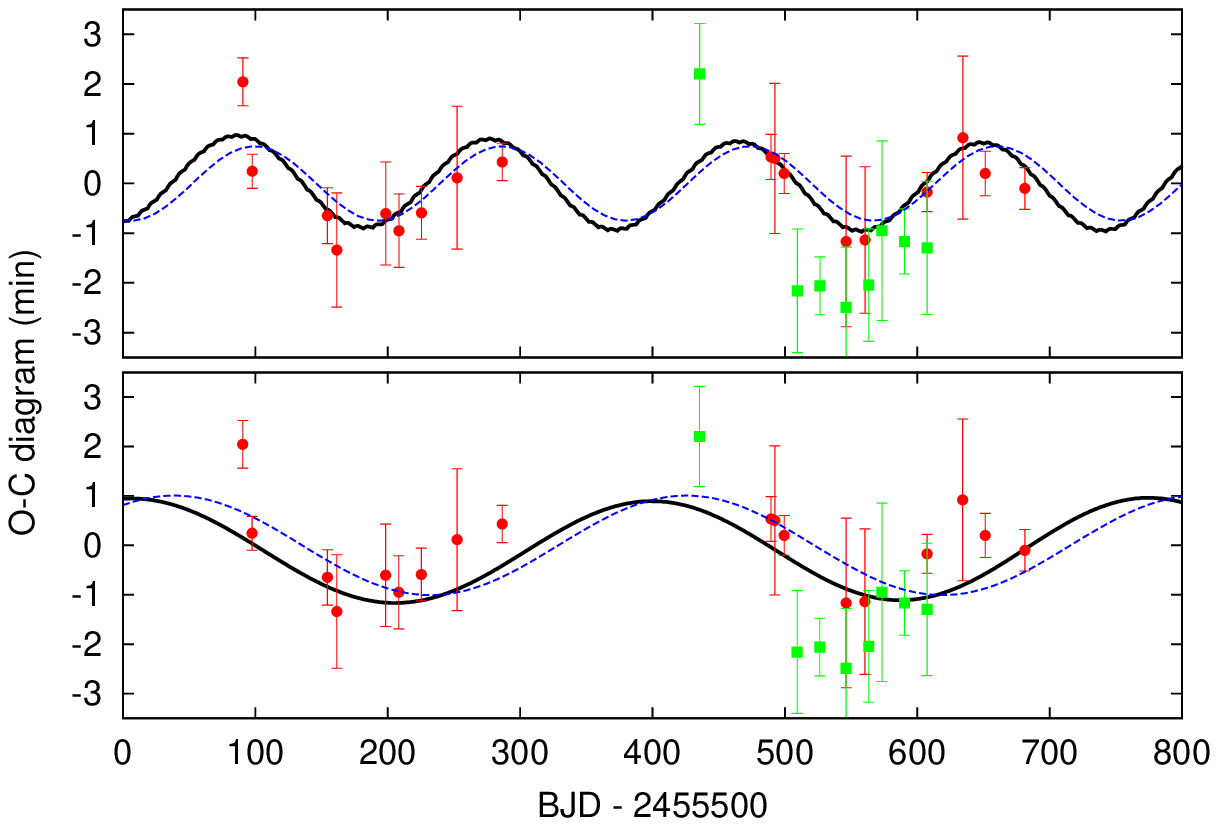}
  \caption{\label{fig:smallPert}OC diagram for OLT (red circles) and
    PTST (green squares) data points plus n-body solutions for two
    different dynamical scenarios (black lines) and our best initial
    sinusoidal fit (blue dashed lines) artificially shifted in phase
    for a better comparison.}
\end{figure}

We thus investigated some possible solutions that would satisfy the
first two TTV periods assuming the specific resonant orbits indicated
in Figure~\ref{fig:scatter}. As an example we show in
Fig.~\ref{fig:smallPert} and Tab.~\ref{tab:smallPert} three arbitrary
cases; similar solutions can be derived with other choices of resonant
orbits. To produce Fig.~\ref{fig:smallPert} we ran our n-body
simulation with specified initial conditions (cf.,
Tab.~\ref{tab:smallPert}) and compared the results with the observed
OC-values and our (original) sinusoidal fit (cf.,
Fig.~\ref{fig:smallPert}). As is clear from Fig.~\ref{fig:smallPert},
all these solutions have $\chi^2$ values similar to our first fitting
attempt. Thus, it is futile to produce a proper fitting procedure.

\begin{table}[ht!]
\caption[]{\label{tab:smallPert}Three possible solutions for our two
  main TTV signals for three different dynamical scenarios. From top
  to bottom, resonances are 5:2, 2:1, and 3:1.}
\centering
\begin{tabular}{l c c}
\hline
\hline
Resonance 5:2, $\sim$190 days &         \qtb      &    Perturber   \\
\hline
Mass ($M_{Jup}$)         &      1.090            &     0.019     \\
Orbital period (days)   &      1.4200246        & 3.550061      \\
Eccentricity            &         0             &     0.3       \\
$\omega$ ($^{\circ}$)     &      274              &   21        \\    
$\Omega$ ($^{\circ}$)     &      235              &    90         \\
$t$                     &          0            &       0       \\
\hline
Resonance 2:1, $\sim$190 days &       \qtb      &    Perturber   \\
\hline
Mass ($M_{Jup}$)         &       1.090            &   0.005           \\ 
Orbital period (days)   &      1.4200246        &  2.8400492          \\
Eccentricity            &          0             &       0.15         \\
$\omega$ ($^{\circ}$)     &      190             &        30          \\
$\Omega$ ($^{\circ}$)     &      330             &       120          \\
$t$                     &         0              &        0           \\
\hline
Resonance 3:1, $\sim$380 days    &     \qtb      &    Perturber   \\
\hline
Mass ($M_{Jup}$)        &      1.090            &     0.035     \\
Orbital period (days)  &      1.4200246        & 4.2600738     \\
Eccentricity           &         0             &     0.135     \\
$\omega$ ($^{\circ}$)    &      202            &      31       \\    
$\Omega$ ($^{\circ}$)    &      300              &    90         \\
$t$                    &          0            &       0       \\
\hline
\end{tabular}
 \tablefoot{$\omega$: longitude of periastron; $\Omega$: longitude of
   ascending node; $t$: time of periastron passage.}
\end{table}

\subsection{Massive perturber in a 190-day orbit}

\begin{figure}[ht!]
  \centering
  \includegraphics[width=0.5\textwidth]{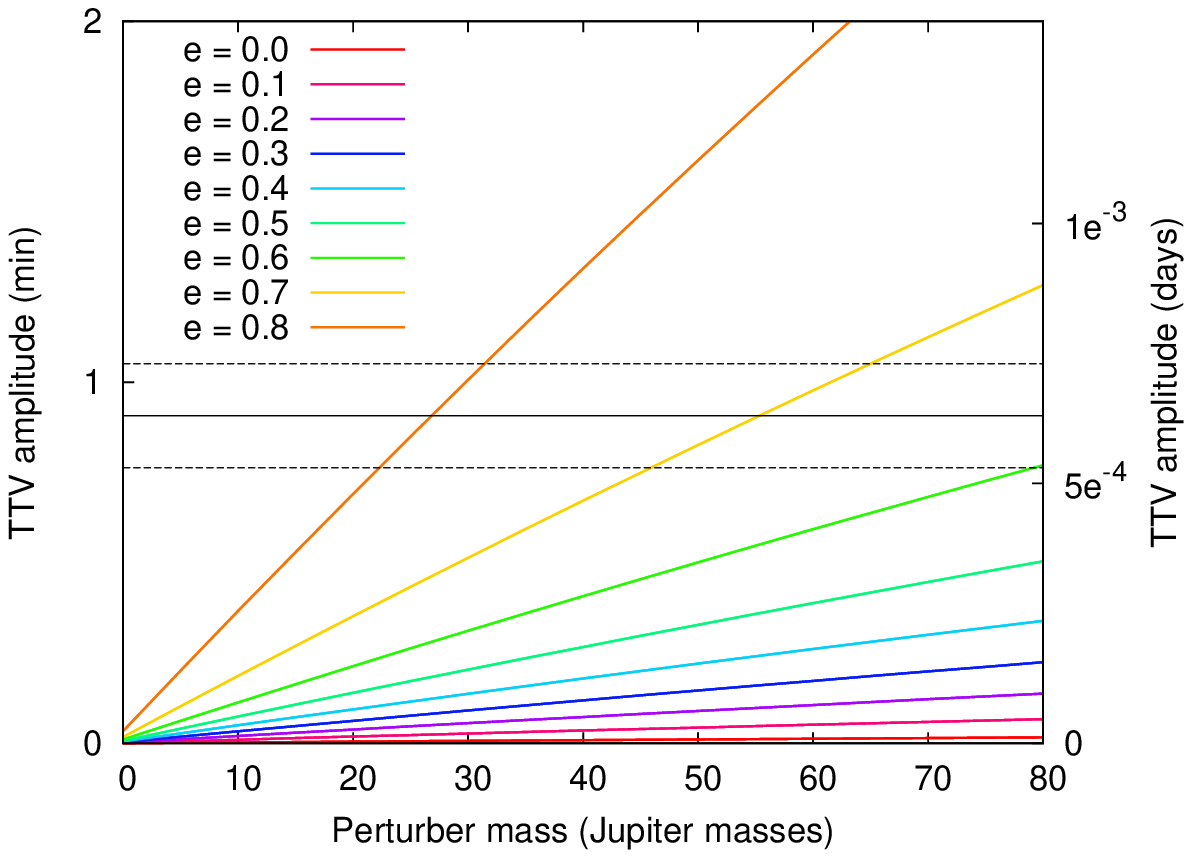}
  \caption{\label{fig:pertmodel} TTV amplitude as a function of the
    mass of the perturber for different eccentricity values. The
    black continuous line shows the best-fitted amplitude to our OC
    diagram, along with 1$\sigma$ errors (dashed lines).}
\end{figure}

As an alternative to a lower-mass perturber in a resonant orbit, we
also considered a more massive perturber in a non-resonant 190-day
orbit. In this case the TTV amplitude strongly depends on the mass and
the eccentricity of the assumed perturber. Again considering the
simplest case of coplanar orbits and both longitude of periastron and
longitude of the ascending node fixed to zero, we computed the
amplitude of the TTV signal as a function of the perturber mass and
its orbital eccentricity using our n-body code, and we compared the
predicted TTV amplitudes with those observed. Our results are shown in
Fig.~\ref{fig:pertmodel}, where we plot the expected TTV amplitude as
a function of the perturber mass for various eccentricities in the range
between zero and 0.8. For low eccentricity values we require
perturber masses of more than 80 Jupiter masses, i.e., we would
require a low-mass star. For eccentricities above 0.6 a brown
dwarf could explain the observed TTV amplitudes.  It is obvious that
such an object would lead to RV variations in the host of Qatar-1,
which could in principle be observed.  Since for a very eccentric
orbit the RV variations are concentrated around the periastron passage
and since only ten days of RV measurements are available for
Qatar-1 so far, any long-term RV variations of Qatar-1 are unknown so far and
could confirm or reject the presence of a massive perturber.

\section{Conclusions}
\label{sec:concl}

Our analysis of the mid-timing residuals of \qt \ taken during almost
two years indicate that the orbital period of the exoplanet is not
constant. The observed long-term timing variations are highly
significant from a statistical point of view and can be explained by
very different physical scenarios.  RV monitoring of \qt \ will
provide an upper limit to the mass of a possible perturber and
continued timing observations of Qatar-1 are required to better
delineate the solution space for the possible perturber geometries.

\begin{acknowledgements}

C. von Essen acknowledges funding by the DFG in the framework of RTG
1351, and P. Hauschildt, S. Witte and H. M. M\"uller for discussions
about the PHOENIX code and limb-darkening effects. 

\end{acknowledgements}

\bibliographystyle{aa}
\bibliography{qatar1ttv}

\begin{thebibliography}{37}
\expandafter\ifx\csname natexlab\endcsname\relax\def\natexlab#1{#1}\fi

\bibitem[{{Adams} {et~al.}(2010){Adams}, {L{\'o}pez-Morales}, {Elliot},
  {Seager}, \& {Osip}}]{Adams2010}
{Adams}, E.~R., {L{\'o}pez-Morales}, M., {Elliot}, J.~L., {Seager}, S., \&
  {Osip}, D.~J. 2010, \apj, 714, 13

\bibitem[{{Agol} {et~al.}(2005){Agol}, {Steffen}, {Sari}, \&
  {Clarkson}}]{Agol2005}
{Agol}, E., {Steffen}, J., {Sari}, R., \& {Clarkson}, W. 2005, \mnras, 359, 567

\bibitem[{{Alsubai} {et~al.}(2011){Alsubai}, {Parley}, {Bramich}, {West},
  {Sorensen}, {Collier Cameron}, {Latham}, {Horne}, {Anderson}, {Bakos},
  {Brown}, {Buchhave}, {Esquerdo}, {Everett}, {F{\.z}r{\'e}sz}, {Hartman},
  {Hellier}, {Miller}, {Pollacco}, {Quinn}, {Smith}, {Stefanik}, \&
  {Szentgyorgyi}}]{Alsubai2011}
{Alsubai}, K.~A., {Parley}, N.~R., {Bramich}, D.~M., {et~al.} 2011, \mnras,
  417, 709

\bibitem[{{Balan} \& {Lahav}(2009)}]{2009MNRAS.394.1936B}
{Balan}, S.~T. \& {Lahav}, O. 2009, \mnras, 394, 1936

\bibitem[{{Ballard} {et~al.}(2011){Ballard}, {Fabrycky}, {Fressin},
  {Charbonneau}, {Desert}, {Torres}, {Marcy}, {Burke}, {Isaacson}, {Henze},
  {Steffen}, {Ciardi}, {Howell}, {Cochran}, {Endl}, {Bryson}, {Rowe}, {Holman},
  {Lissauer}, {Jenkins}, {Still}, {Ford}, {Christiansen}, {Middour}, {Haas},
  {Li}, {Hall}, {McCauliff}, {Batalha}, {Koch}, \&
  {Borucki}}]{2011ApJ...743..200B}
{Ballard}, S., {Fabrycky}, D., {Fressin}, F., {et~al.} 2011, \apj, 743, 200

\bibitem[{{Carter} \& {Winn}(2009)}]{Carter2009}
{Carter}, J.~A. \& {Winn}, J.~N. 2009, \apj, 704, 51

\bibitem[{{Claret} \& {Hauschildt}(2003)}]{ClaretHauschildt2003}
{Claret}, A. \& {Hauschildt}, P.~H. 2003, \aap, 412, 241

\bibitem[{{Cochran} {et~al.}(2011){Cochran}, {Fabrycky}, {Torres}, {Fressin},
  {D{\'e}sert}, {Ragozzine}, {Sasselov}, {Fortney}, {Rowe}, {Brugamyer},
  {Bryson}, {Carter}, {Ciardi}, {Howell}, {Steffen}, {Borucki}, {Koch}, {Winn},
  {Welsh}, {Uddin}, {Tenenbaum}, {Still}, {Seager}, {Quinn}, {Mullally},
  {Miller}, {Marcy}, {MacQueen}, {Lucas}, {Lissauer}, {Latham}, {Knutson},
  {Kinemuchi}, {Johnson}, {Jenkins}, {Isaacson}, {Howard}, {Horch}, {Holman},
  {Henze}, {Haas}, {Gilliland}, {Gautier}, {Ford}, {Fischer}, {Everett},
  {Endl}, {Demory}, {Deming}, {Charbonneau}, {Caldwell}, {Buchhave}, {Brown},
  \& {Batalha}}]{Cochran2011}
{Cochran}, W.~D., {Fabrycky}, D.~C., {Torres}, G., {et~al.} 2011, \apjs, 197, 7

\bibitem[{{Covino} {et~al.}(2013){Covino}, {Esposito}, {Barbieri}, {Mancini},
  {Nascimbeni}, {Claudi}, {Desidera}, {Gratton}, {Lanza}, {Sozzetti}, {Biazzo},
  {Affer}, {Gandolfi}, {Munari}, {Pagano}, {Bonomo}, {Collier Cameron},
  {H{\'e}brard}, {Maggio}, {Messina}, {Micela}, {Molinari}, {Pepe}, {Piotto},
  {Ribas}, {Santos}, {Southworth}, {Shkolnik}, {Triaud}, {Bedin}, {Benatti},
  {Boccato}, {Bonavita}, {Borsa}, {Borsato}, {Brown}, {Carolo}, {Ciceri},
  {Cosentino}, {Damasso}, {Faedi}, {Mart{\'{\i}}nez Fiorenzano}, {Latham},
  {Lovis}, {Mordasini}, {Nikolov}, {Poretti}, {Rainer}, {Rebolo L{\'o}pez},
  {Scandariato}, {Silvotti}, {Smareglia}, {Alcala}, {Cunial}, {Di Fabrizio},
  {Di Mauro}, {Giacobbe}, {Granata}, {Harutyunyan}, {Knapic}, {Lattanzi},
  {Leto}, {Lodato}, {Malavolta}, {Marzari}, {Molinaro}, {Nardiello}, {Pedani},
  {Prisinzano}, \& {Turrini}}]{Covino2013}
{Covino}, E., {Esposito}, M., {Barbieri}, M., {et~al.} 2013, ArXiv e-prints

\bibitem[{{D{\'{\i}}az} {et~al.}(2008){D{\'{\i}}az}, {Rojo}, {Melita}, {Hoyer},
  {Minniti}, {Mauas}, \& {Ru{\'{\i}}z}}]{Diaz2008}
{D{\'{\i}}az}, R.~F., {Rojo}, P., {Melita}, M., {et~al.} 2008, \apjl, 682, L49

\bibitem[{{Eastman} {et~al.}(2010){Eastman}, {Siverd}, \&
  {Gaudi}}]{Eastman2010}
{Eastman}, J., {Siverd}, R., \& {Gaudi}, B.~S. 2010, \pasp, 122, 935

\bibitem[{{Fabrycky} {et~al.}(2012){Fabrycky}, {Ford}, {Steffen}, {Rowe},
  {Carter}, {Moorhead}, {Batalha}, {Borucki}, {Bryson}, {Buchhave},
  {Christiansen}, {Ciardi}, {Cochran}, {Endl}, {Fanelli}, {Fischer}, {Fressin},
  {Geary}, {Haas}, {Hall}, {Holman}, {Jenkins}, {Koch}, {Latham}, {Li},
  {Lissauer}, {Lucas}, {Marcy}, {Mazeh}, {McCauliff}, {Quinn}, {Ragozzine},
  {Sasselov}, \& {Shporer}}]{Fabrycky2012}
{Fabrycky}, D.~C., {Ford}, E.~B., {Steffen}, J.~H., {et~al.} 2012, \apj, 750,
  114

\bibitem[{{Fukui} {et~al.}(2011){Fukui}, {Narita}, {Tristram}, {Sumi}, {Abe},
  {Itow}, {Sullivan}, {Bond}, {Hirano}, {Tamura}, {Bennett}, {Furusawa},
  {Hayashi}, {Hearnshaw}, {Hosaka}, {Kamiya}, {Kobara}, {Korpela}, {Kilmartin},
  {Lin}, {Ling}, {Makita}, {Masuda}, {Matsubara}, {Miyake}, {Muraki}, {Nagaya},
  {Nishimoto}, {Ohnishi}, {Omori}, {Perrott}, {Rattenbury}, {Saito}, {Skuljan},
  {Suzuki}, {Sweatman}, \& {Wada}}]{Fukui2011}
{Fukui}, A., {Narita}, N., {Tristram}, P.~J., {et~al.} 2011, \pasj, 63, 287

\bibitem[{{Fulton} {et~al.}(2011){Fulton}, {Shporer}, {Winn}, {Holman},
  {P{\'a}l}, \& {Gazak}}]{Fulton2011}
{Fulton}, B.~J., {Shporer}, A., {Winn}, J.~N., {et~al.} 2011, \aj, 142, 84

\bibitem[{{Haario} {et~al.}(2001){Haario}, {Saksman}, \&
  {Tamminen}}]{Haario2001}
{Haario}, H., {Saksman}, E., \& {Tamminen}, J. 2001, Bernoulli, 7 (2), 223

\bibitem[{Hauschildt \& Baron(1999)}]{Peter1}
Hauschildt, P.~H. \& Baron, E. 1999, Journal of Computational and Applied
  Mathematics, 109, 41

\bibitem[{{Holman} {et~al.}(2010){Holman}, {Fabrycky}, {Ragozzine}, {Ford},
  {Steffen}, {Welsh}, {Lissauer}, {Latham}, {Marcy}, {Walkowicz}, {Batalha},
  {Jenkins}, {Rowe}, {Cochran}, {Fressin}, {Torres}, {Buchhave}, {Sasselov},
  {Borucki}, {Koch}, {Basri}, {Brown}, {Caldwell}, {Charbonneau}, {Dunham},
  {Gautier}, {Geary}, {Gilliland}, {Haas}, {Howell}, {Ciardi}, {Endl},
  {Fischer}, {F{\"u}r{\'e}sz}, {Hartman}, {Isaacson}, {Johnson}, {MacQueen},
  {Moorhead}, {Morehead}, \& {Orosz}}]{Holman2010}
{Holman}, M.~J., {Fabrycky}, D.~C., {Ragozzine}, D., {et~al.} 2010, Science,
  330, 51

\bibitem[{{Holman} \& {Murray}(2005)}]{Holman2005}
{Holman}, M.~J. \& {Murray}, N.~W. 2005, Science, 307, 1288

\bibitem[{{Irwin} {et~al.}(2010){Irwin}, {Buchhave}, {Berta}, {Charbonneau},
  {Latham}, {Burke}, {Esquerdo}, {Everett}, {Holman}, {Nutzman}, {Berlind},
  {Calkins}, {Falco}, {Winn}, {Johnson}, \& {Gazak}}]{2010ApJ...718.1353I}
{Irwin}, J., {Buchhave}, L., {Berta}, Z.~K., {et~al.} 2010, \apj, 718, 1353

\bibitem[{{Irwin} {et~al.}(2011){Irwin}, {Quinn}, {Berta}, {Latham}, {Torres},
  {Burke}, {Charbonneau}, {Dittmann}, {Esquerdo}, {Stefanik}, {Oksanen},
  {Buchhave}, {Nutzman}, {Berlind}, {Calkins}, \&
  {Falco}}]{2011ApJ...742..123I}
{Irwin}, J.~M., {Quinn}, S.~N., {Berta}, Z.~K., {et~al.} 2011, \apj, 742, 123

\bibitem[{{Jones} {et~al.}(2001){Jones}, {Oliphant}, {Peterson},
  {et~al.}}]{Jones2001}
{Jones}, E., {Oliphant}, T., {Peterson}, P., {et~al.} 2001, {SciPy}: Open
  source scientific tools for {Python}, \url{http://www.scipy.org}

\bibitem[{{Kipping}(2010)}]{Kipping2010}
{Kipping}, D.~M. 2010, \mnras, 408, 1758

\bibitem[{{Lissauer} {et~al.}(2011){Lissauer}, {Fabrycky}, {Ford}, {Borucki},
  {Fressin}, {Marcy}, {Orosz}, {Rowe}, {Torres}, {Welsh}, {Batalha}, {Bryson},
  {Buchhave}, {Caldwell}, {Carter}, {Charbonneau}, {Christiansen}, {Cochran},
  {Desert}, {Dunham}, {Fanelli}, {Fortney}, {Gautier}, {Geary}, {Gilliland},
  {Haas}, {Hall}, {Holman}, {Koch}, {Latham}, {Lopez}, {McCauliff}, {Miller},
  {Morehead}, {Quintana}, {Ragozzine}, {Sasselov}, {Short}, \&
  {Steffen}}]{Lissauer2011}
{Lissauer}, J.~J., {Fabrycky}, D.~C., {Ford}, E.~B., {et~al.} 2011, \nat, 470,
  53

\bibitem[{{Lomb}(1976)}]{Lomb}
{Lomb}, N.~R. 1976, \apss, 39, 447

\bibitem[{{Maciejewski} {et~al.}(2010){Maciejewski}, {Dimitrov},
  {Neuh{\"a}user}, {Niedzielski}, {Raetz}, {Ginski}, {Adam}, {Marka},
  {Moualla}, \& {Mugrauer}}]{Maciejewski2010}
{Maciejewski}, G., {Dimitrov}, D., {Neuh{\"a}user}, R., {et~al.} 2010, \mnras,
  407, 2625

\bibitem[{{Maciejewski} {et~al.}(2011){Maciejewski}, {Dimitrov},
  {Neuh{\"a}user}, {Tetzlaff}, {Niedzielski}, {Raetz}, {Chen}, {Walter},
  {Marka}, {Baar}, {Krejcov{\'a}}, {Budaj}, {Krushevska}, {Tachihara},
  {Takahashi}, \& {Mugrauer}}]{Maciejewski2011}
{Maciejewski}, G., {Dimitrov}, D., {Neuh{\"a}user}, R., {et~al.} 2011, \mnras,
  411, 1204

\bibitem[{{Mandel} \& {Agol}(2002)}]{MandelAgol2002}
{Mandel}, K. \& {Agol}, E. 2002, \apjl, 580, L171

\bibitem[{{Nascimbeni} {et~al.}(2011){Nascimbeni}, {Piotto}, {Bedin},
  {Damasso}, {Malavolta}, \& {Borsato}}]{Nascimbeni2011}
{Nascimbeni}, V., {Piotto}, G., {Bedin}, L.~R., {et~al.} 2011, \aap, 532, A24

\bibitem[{{P{\'a}l} {et~al.}(2011){P{\'a}l}, {S{\'a}rneczky}, {Szab{\'o}},
  {Szing}, {Kiss}, {Mez{\H o}}, \& {Reg{\'a}ly}}]{Pal2011}
{P{\'a}l}, A., {S{\'a}rneczky}, K., {Szab{\'o}}, G.~M., {et~al.} 2011, \mnras,
  413, L43

\bibitem[{{Patil} {et~al.}(2010){Patil}, {Huard}, \& {Fonnesbeck}}]{Patil2010}
{Patil}, A., {Huard}, D., \& {Fonnesbeck}, C.~J. 2010, Journal of Statistical
  Software, 35, 1

\bibitem[{{Poddan{\'y}} {et~al.}(2010){Poddan{\'y}}, {Br{\'a}t}, \&
  {Pejcha}}]{ETD}
{Poddan{\'y}}, S., {Br{\'a}t}, L., \& {Pejcha}, O. 2010, \na, 15, 297

\bibitem[{{Pont} {et~al.}(2006){Pont}, {Zucker}, \& {Queloz}}]{Pont2006}
{Pont}, F., {Zucker}, S., \& {Queloz}, D. 2006, \mnras, 373, 231

\bibitem[{{Scargle}(1982)}]{Scargle}
{Scargle}, J.~D. 1982, \apj, 263, 835

\bibitem[{{Steffen} {et~al.}(2012){Steffen}, {Ragozzine}, {Fabrycky}, {Carter},
  {Ford}, {Holman}, {Rowe}, {Welsh}, {Borucki}, {Boss}, {Ciardi}, \&
  {Quinn}}]{Steffen2012}
{Steffen}, J.~H., {Ragozzine}, D., {Fabrycky}, D.~C., {et~al.} 2012,
  Proceedings of the National Academy of Science, 109, 7982

\bibitem[{{Vihola}(2011)}]{Vihola2011}
{Vihola}, M. 2011, {Stochastic Processes and their Applications}, {in press,
  ArXiv e-prints: 0903.4061}

\bibitem[{{Witte} {et~al.}(2009){Witte}, {Helling}, \& {Hauschildt}}]{Peter2}
{Witte}, S., {Helling}, C., \& {Hauschildt}, P.~H. 2009, \aap, 506, 1367

\bibitem[{{Zechmeister} \& {K{\"u}rster}(2009)}]{LombScargle}
{Zechmeister}, M. \& {K{\"u}rster}, M. 2009, \aap, 496, 577

\end{thebibliography}

\end{document}